\documentclass[12pt]{article}

\usepackage{amsmath}
\usepackage{newtxtext}
\usepackage[subscriptcorrection]{newtxmath}
\usepackage{bm}
\usepackage{comment}
\usepackage{graphicx}
\usepackage[plain,noend]{algorithm2e}
\usepackage{apacite}  % APA citation style package
\usepackage{geometry} % マージン調整パッケージ
\usepackage{algorithmic}
\usepackage{natbib}
\usepackage{authblk}
\usepackage{multirow}
\usepackage[colorlinks=true, linkcolor=blue, citecolor=blue, urlcolor=blue]{hyperref}

\newtheorem{theorem}{Theorem}
\newtheorem{corollary}{Corollary}

\makeatletter
\renewcommand{\algocf@captiontext}[2]{#1\algocf@typo. \AlCapFnt{}#2} % text of caption
\def\@algocf@capt@plain{top}
\renewcommand{\algocf@makecaption}[2]{%
  \addtolength{\hsize}{\algomargin}%
  \sbox\@tempboxa{\algocf@captiontext{#1}{#2}}%
  \ifdim\wd\@tempboxa >\hsize%     % if caption is longer than a line
    \hskip .5\algomargin%
    \parbox[t]{\hsize}{\algocf@captiontext{#1}{#2}}% then caption is not centered
  \else%
    \global\@minipagefalse%
    \hbox to\hsize{\box\@tempboxa}% else caption is centered
  \fi%
  \addtolength{\hsize}{-\algomargin}%
}
\makeatother

\begin{document}

\title{\bf %Discrete divergence approximations to the latent correlation coefficient in two-Way contingency tables
A family of divergence-based correlation measures for contingency tables under bivariate normality}

\author[1]{Wataru Urasaki}
%\author[2]{Tomoyuki Nakagawa}
%\author[1]{Kouji Tahata}

\affil[1]{Department of Information Science and Technology, Tokyo University of Science}
%\affil[2]{School of Data Science, Meisei University}
\date{}

\begin{comment}
\author{W. URASAKI \\
Department of Information Sciences, Tokyo University of Science, \\ Noda City, 278-8510, Chiba, Japan \\
\texttt{urasaki.stat@gmail.com} \and
G. KAWAMITSU \\
Department of Information Sciences, Tokyo University of Science, \\ Noda City, 278-8510, Chiba, Japan \\
\texttt{6322513@ed.tus.ac.jp} \and
T. NAKAGAWA \\
School of Data Science, Meisei University,\\ Hino City, 191-8506, Tokyo, Japan \\
\texttt{tomoyuki.nakagawa@meisei-u.ac.jp} \and
K. TAHATA \\
Department of Information Sciences, Tokyo University of Science, \\ Noda City, 278-8510, Chiba, Japan \\
\texttt{kouji\_tahata@is.noda.tus.ac.jp}}
\end{comment}

\maketitle

\begin{abstract}
We propose a family of association measures for two-way contingency tables whose latent distribution can be assumed to be bivariate normal.
When this assumption holds, the power-divergence measuring departure from independence can be approximated in closed form as a function of the latent correlation coefficient.
By inverting this relationship, we obtain a family of measures $\rho_{(\lambda)}$, indexed by a scalar parameter $-1 \leq \lambda \leq 1$, that directly approximates the latent correlation.
Special cases include the informational measure of correlation proposed by Linfoot (1957) at $\lambda = 0$ and Pearson's contingency coefficient $C$ at $\lambda = 1$.
Additionally, we derive asymptotic distributions via the delta method and construct two families of confidence intervals.
Simulation studies confirm that the proposed measures approximate the true latent correlation more faithfully than conventional divergence-based measures, and that they successfully distinguish between weak and moderate associations where existing measures tend to give indistinguishable values.
Compared with the polychoric correlation coefficient, the proposed measures are computed several thousand times faster and remain numerically stable even when the latent correlation is close to one.
%We recommend using $\lambda = 0$, $2/3$, and $1$ in practice.
\end{abstract}

\medskip


\medskip


%%%%%%%%%%%%%%%%%%%%%%%%%%%%%%%%%%%%%%%%%%%%%%%%%%%%%%%%%%%%%%%%%%%%%%%%%%%%
\section{Introduction}
Categorical variables, which originate from distinct categories, are widely observed in medicine, education, psychology, and social science, and their statistical analysis has attracted considerable attention for more than a century.
When two categorical variables are observed together, their joint distribution is naturally summarized in a two-way contingency table.
A central question in the analysis of such a table is whether the row and column variables are statistically independent.
Pearson's chi-squared test \citep{pearson1900} and the log-likelihood ratio test \citep{wilks1935likelihood} are the most prominent procedures for testing independence, and both are implemented as standard routines in widely used statistical software such as R and SAS.
When independence is rejected, or when the variables are clearly related, researchers commonly turn to measures of association to quantify the strength of the relationship.

Various association measures have been proposed for two-way contingency tables.
Well-known examples include Cram\'{e}r's coefficient $V^{2}$ \citep{cramer1946mathematical} and Theil's uncertainty coefficient \citep{theil1970estimation}, both of which quantify the degree of departure from independence within a bounded interval and are unaffected by sample size.
More recently, \citet{urasaki2024generalized} proposed a generalized family
of such measures by linking the contingency table to the generalized divergence, unifying and extending earlier divergence-based proposals \citep{kateri1994f, kateri1997asymmetry, forcina2021new, tomizawa2004generalization}.
However, a common limitation of these measures is that their values do not have a natural connection to the Pearson correlation coefficient, which is the most familiar and widely understood index of association for continuous data.
As a result, it is often difficult for practitioners to judge in concrete terms how strong an association is, even after computing one of these measures.

A different perspective on the problem is provided by the latent variable approach to contingency table analysis.
The idea, which can be traced back to \citet{pearson1900}, is that observed categorical responses are often the result of discretizing inherently continuous underlying variables, such as ability, attitude, or severity, at threshold values.
If a bivariate normal distribution is assumed for these latent variables, then the appropriate target of inference is the correlation coefficient between them.
This is precisely the quantity estimated by the polychoric correlation coefficient (\citealp{olsson1979maximum}; see also, \citealp{agresti2010analysis}), which has become a standard tool in psychometrics and structural equation modeling.
The polychoric correlation is directly comparable to the Pearson correlation, making it highly interpretable in practice.
However, its maximum likelihood estimator requires numerical optimization over a two-dimensional integral, which is computationally expensive when the number of categories is large.
Furthermore, the estimator is known to become numerically unstable when the true latent correlation is close to one, which is precisely the situation that arises in practice when two variables are strongly related, and in such cases the procedure may fail to converge or may produce estimates at the boundary of the parameter space.
These drawbacks make it difficult to apply the polychoric correlation routinely in large-scale exploratory analyses, such as the simultaneous screening of many variable pairs from a questionnaire survey.

In this paper, we propose a new family of association measures that inherits the interpretability of the polychoric correlation coefficient while avoiding its computational and numerical difficulties.
Our starting point is a key result of \citet{urasaki2024generalized}, which showed that, when the latent distribution of a two-way contingency table is bivariate normal, the $f$-divergence measuring departure from independence can be approximated in closed form as a function of the latent correlation coefficient alone.
In particular, within the power-divergence family \citep{cressie1984multinomial}, parameterized by a scalar $-1\leq \lambda \leq 1$, this approximation expresses the computable divergence as a simple algebraic function of the squared latent correlation.
By inverting this relationship, one can recover an approximation to the latent correlation from the observed divergence, without computing any numerical integrals.
The resulting family of measures, which we denote by $\rho_{(\lambda)}$, can be evaluated orders of magnitude faster than the polychoric correlation coefficient and remains numerically stable even when the underlying association is strong.

Several connections to existing measures arise naturally within this framework.
When $\lambda = 0$, corresponding to the Kullback--Leibler divergence, the proposed measure coincides with the informational measure of correlation introduced by \citet{linfoot1957informational} and, from a model-selection perspective, with Cox--Snell's pseudo $R^{2}$ \citep{cox1989analysis} for the multinomial logit model.
When $\lambda = 1$, corresponding to the Pearson divergence, it coincides with Pearson's contingency coefficient $C$ \citep{pearson1904theory}, which was itself proposed over a century ago as an approximation to a latent normal correlation.
These connections reveal that the proposed framework provides a principled generalization of classical results, extending them to a continuous family of divergence-based correlation measures indexed by $\lambda$.
In addition to point estimation, we develop a complete inferential framework for practical use.
We derive asymptotic distributions for $\rho_{(\lambda)}$ and $\rho^{2}_{(\lambda)}$ via the delta method and construct two families of confidence intervals: a direct interval based on the asymptotic normality of the estimator (Simple method), and one based on Fisher's $z$-transformation \citep[Fisher's $z$ method; see also][]{itaya2025asymptotic}, which is known to perform well for correlation coefficients.
We also introduce detectable correlation thresholds that allow practitioners to assess not merely whether an association is statistically significant, but whether its magnitude is practically meaningful given the sample size and significance level.

%Through simulation experiments, we confirm that the proposed measures approximate the true latent correlation more faithfully than conventional divergence-based measures such as Cram\'{e}r's $V^{2}$ and the total uncertainty coefficient \citep{momozaki2023extension}, and that they can successfully distinguish between weak and moderate associations where existing measures tend to give indistinguishable values.
%In a direct comparison with the polychoric correlation coefficient, the proposed measures are computed several thousand times faster for large tables and remain stable even when the latent correlation approaches one, a regime where the polychoric estimator is known to deteriorate.
%Based on these findings, we recommend using the parameter values $\lambda = 0$, $2/3$, and $1$, corresponding to the Kullback--Leibler, Cressie--Read \citep{cressie1984multinomial}, and Pearson divergences, as the most practical choices for applied data analysis.

The remainder of the paper is organized as follows. 
Section~2 introduces the latent distribution framework for two-way contingency tables and the generalized divergence, and derives the approximation connecting the power-divergence to the latent correlation coefficient under the bivariate normal assumption.
Special cases and their relationships to existing measures are also discussed in that section.
Section~3 presents the proposed association measures, a Newton's method algorithm for their computation, and the two types of confidence intervals.
Section~4 reports simulation studies evaluating the accuracy of the measures and the coverage of the confidence intervals.
Section~5 applies the proposed measures to real data from a national health survey to illustrate their practical use.
Section~6 provides concluding remarks.

%%%%%%%%%%%%%%%%%%%%%%%%%%%%%%%%%%%%%%%%%%%%%%%%%%%%%%%%%%%%%%%%%%%%%%%%%%%%
\section{Latent distributions and divergence framework in two-way contingency tables}
Consider an $r\times c$ contingency table with ordinal categories for the row variable $X$ and the column variable $Y$.
Let $p_{ij}$ denote the probability that an observation will fall in the $i$th row and $j$th column of the table ($i=1,\ldots, r;j=1,\ldots, c)$.
Let $p_{i\cdot}=\sum_{j=1}^c p_{ij}$ and $p_{\cdot j}=\sum_{i=1}^r p_{ij}$ denote the marginal probabilities. 
Conversely, let $n_{ij}$ denote the observed frequency in the $i$th row and $j$th column of the table.
The totals $n_{i\cdot}$, $n_{\cdot j}$ and $n$ are also denoted as $n_{i\cdot}=\sum_{j=1}^c n_{ij}$, $n_{\cdot j}=\sum_{i=1}^r n_{ij}$, and $n = \sum_{i=1}^r\sum_{j=1}^c n_{ij}$, respectively.
Furthermore, when analyzing the real data, it is common to assume a multinomial distribution for the $r\times c$ contingency table and to replace the cell probabilities $p_{ij}$ with their estimates $\hat{p}_{ij}=n_{ij}/n$.
Using this notation, we introduce latent distributions and divergence framework in two-way contingency tables.

%%%%%%%%%%%%%%%%%%%%%%%%%%%%%%%%%%%%%%%%%%%%%%%%%%%%%%
\subsection{Generalized divergence in two-way contingency tables}\label{Sec2.2}
For the $r \times c$ contingency table, $P$ and $Q$ are given as discrete distributions $\{p_{ij}\}$ and $\{q_{ij}\}$. 
Accordingly, we have $dP/dQ = \{p_{ij}/q_{ij}\}$. 
As a measure of these distributions, the $f$-divergence from $\{p_{ij}\}$ to $\{q_{ij}\}$ is given as
\begin{align*}
I_f(P;Q) = I_f(\{p_{ij}\};\{q_{ij}\}) &= \sum^r_{i=1}\sum^c_{j=1} q_{ij} f\left(\frac{p_{ij}}{ q_{ij}}  \right), 
\end{align*}
where $f(x)$ is a once-differentiable and strictly convex function on $(0, +\infty)$ with $f(1) = 0$, $ \lim_{x \to 0}f(x) = 0$, $0f(0/0) = 0$, and $0f(a/0) = a\lim_{x \to \infty}f(x) / x$ (see, \citealp{csiszar2004information}). 
By choosing function $f$, many important divergences, such as the Kullback–Leibler (KL) divergence,  ($f(x) = x\log x$), the Pearson divergence ($f(x) = x^2-x$), the power-divergence ($f(x)=(x^{\lambda+1}-x)/\lambda(\lambda+1)$, $(-\infty < \lambda < \infty)$), and the $\theta$-divergence ($f(x) = (x-1)^2/(\theta x + 1 - \theta) + (x-1)/(1 - \theta)$, $(0 \leq \theta \leq 1)$), are included in special cases of the $f$-divergence (see, e.g., \citealp{Sason2016divergence, ichimori2013inequalities}).
In particular, power and $\theta$-divergences are also generalized divergences that include several important divergences, but the $f$-divergence is a broad class of generalized divergences that further includes these.
Furthermore, the $f$-divergence is one of the monotone and regular divergences. 
The class of monotone and regular divergences is introduced in \cite{cencov2000statistical} and studied in \cite{corcuera1998characterization} as a wide class of invariant divergences with respect to Markov embeddings. 
The class of monotone and regular divergence is often used as a measure of goodness of prediction (see, e.g., \citealp{gkisser1993predictive, corcuera1999relationship, corcuera1999generalized}).

One well-known application of divergence in the analysis of two-way contingency tables, even among those not familiar with mathematical statistics, is the test of independence between two variables. 
The test of independence uses the divergence
\begin{align*}
I_f(\{p_{ij}\};\{p_{i\cdot} p_{\cdot j}\}) &= \sum^r_{i=1}\sum^c_{j=1} p_{i\cdot} p_{\cdot j} f\left(\frac{p_{ij}}{p_{i\cdot} p_{\cdot j}}  \right)
\end{align*}
between the two distributions {$p_{ij}$} and {$p_{i\cdot}p_{\cdot j}$}. 
In particular, when $I_f(\{p_{ij}\};\{p_{i\cdot} p_{\cdot j}\})$ is replaced by its plug in estimator $I_f(\{\hat{p}_{ij}\};\{\hat{p}_{i\cdot} \hat{p}_{\cdot j}\})$ obtained by substituting $\hat p_{ij}$ for $p_{ij}$, the statistic $2nI_f(\{\hat{p}_{ij}\};\{\hat{p}_{i\cdot} \hat{p}_{\cdot j}\})$ asymptotically follows a chi-squared distribution with $(r-1)(c-1)$ degrees of freedom. 
Based on this fact, independence is assessed at the $\alpha \%$ significance level.
Although the description here is given in a generalized form, in the cases of the Pearson divergence or the KL divergence, the resulting statistics coincide with standard ones such as Pearson's chi-squared test statistic (\citealp{pearson1900}) and the log likelihood ratio test statistic (\citealp{wilks1935likelihood}), both of which are implemented in statistical software such as R and SAS.

In addition to independence, considerable literature has developed around modeling and goodness-of-fit testing with $f$-divergence in contingency table analysis \citep[e.g.,][etc.]{kateri1994f, kateri1997asymmetry, kateri2007class, forcina2021new, tahata2022advances, pardo2018statistical, felipe2014phi, felipe2018statistical}.

%%%%%%%%%%%%%%%%%%%%%%%%%%%%%%%%%%%%%%%%%%%%%%%%%%%%%%
\subsection{Relationship between latent distributions and generalized divergences}
When a latent distribution is assumed for the probabilistic structure of a two-way contingency table, we follow the paper by \cite{urasaki2024generalized} in describing the relationship between the $f$-divergence for independence and the latent distribution.

Assuming a latent variable, the ($i$,$j$) cell probability $p_{ij}$ of the $r \times c$ contingency table is denoted as
\begin{align*}
p_{ij} &= P(X = i, Y=j) \\
&= P(x_{i-1} < X^* \leq x_i, y_{j-1} < Y^* \leq y_j) \\
&= f_{X^*, Y^*}(\tilde{x}_i, \tilde{y}_j)\Delta_{x_i} \Delta_{y_j},
\end{align*}
where $x_{i-1} < \tilde{x}_i \leq x_i$, $y_{j-1} < \tilde{y}_j \leq y_j$ and  $f_{X^*, Y^*}(\tilde{x}_i, \tilde{y}_j)$ is a continuous joint density function of random variables $X^*$ and $Y^*$.
$\Delta_{x_i}$ and $\Delta_{y_j}$ are the width of intervals $(x_{i-1}, x_{i}]$ and $(y_{j-1}, y_{j}]$, respectively. 
In this situation, it is possible to approximate $I_f(\{p_{ij}\};\{p_{i \cdot}p_{\cdot j}\})$ as follows:
\begin{align*}
\begin{split}
I_f(\{p_{ij}\};\{p_{i \cdot}p_{\cdot j}\}) &= \sum^r_{i=1} \sum^c_{j=1} f_{X^*}(\tilde{x}_i) f_{Y^*}(\tilde{y}_j) f\left(\frac{f_{X^*,Y^*}(\tilde{x}_i, \tilde{y}_j)}{f_{X^*}(\tilde{x}_i) f_{Y^*}(\tilde{y}_j)} \right)\Delta_{x_i} \Delta_{y_j} \\
&\xrightarrow[\Delta_{x_i} \Delta_{y_j} \to 0]{} \int^{\infty}_{-\infty} \int^{\infty}_{-\infty} f_{X^*}(x) f_{Y^*}(y) f\left(\frac{f_{X^*,Y^*}(x, y)}{f_{X^*}(x) f_{Y^*}(y)} \right) dx dy,
\end{split}
\end{align*}
where $f_{X^*}(x)$ and $f_{Y^*}(y)$ are marginal probability density functions of $f_{X^*,Y^*}(x, y)$.

Let $X^*$ and $Y^*$ be random variables according to the bivariate normal distribution and the joint density function is
\begin{align*}
\begin{split}
f_{X^*,Y^*}(x, y) &= \frac{1}{2\pi \sigma_x \sigma_y \sqrt{1-\rho^2}} \exp \left[ -\frac{1}{2(1-\rho^2)} \right. \\
&\quad \left. \qquad \left\lbrace \left(\frac{x-\mu_x}{\sigma_x} \right)^2 - 2\rho \left(\frac{x-\mu_x}{\sigma_x}  \right) \left(\frac{y-\mu_y}{\sigma_y}  \right) + \left(\frac{y-\mu_y}{\sigma_y}  \right)^2 \right\rbrace \right] \\
&\quad -\infty < x < +\infty, \quad -\infty < y < +\infty 
\end{split} 
\end{align*}
where $\rho$ is the correlation coefficient between $X^*$ and $Y^*$. 
The value of the correlation coefficient ranges from $-1$ to $1$. 
In the formula, the standard deviations $\sigma_x$ and $\sigma_y$ are required to be positive, whereas the means $\mu_x$ and $\mu_y$ do not may take any real values.
When applying $f(x)=(x^{\lambda+1}-x)/\lambda(\lambda+1)$, the relationship between the power-divergence and the correlation coefficient $\rho$ is expressed as
\begin{align*}
I_\lambda(\{p_{ij}\};\{p_{i \cdot}p_{\cdot j}\}) &\approx \frac{1}{\lambda(\lambda + 1)} \left\lbrace (1-\rho^2)^{-\frac{\lambda}{2}}(1-\lambda^2 \rho^2)^{-\frac{1}{2}}-1 \right\rbrace,
\end{align*}
where, $I_\lambda(\cdot;\cdot)$ is called a power-divergence and defined as
\begin{align*}
I_\lambda(\{p_{ij}\};\{p_{i \cdot}p_{\cdot j}\}) &= \frac{1}{\lambda(\lambda+1)}\sum^r_{i=1}\sum^c_{j=1} p_{ij} \left\{\left(\frac{p_{ij}}{ p_{i \cdot}p_{\cdot j}}\right)^\lambda - 1\right\}, \quad -\infty < \lambda < +\infty.
\end{align*}
Then, as a condition for integrability we obtain
\begin{align*}
\vert \lambda\rho\vert < 1
\end{align*}
Given the integrability condition $\vert \lambda\rho\vert < 1$ and the range of $\rho$, it follows that $\lambda$ is restricted to the interval $-1\leq \lambda \leq 1$.

%%%%%%%%%%%%%%%%%%%%%%%%%%%%%%%%%%%%%%%%%%%%%%%%%%%%%%
\subsection{Special cases: KL and Pearson divergences}\label{Sec2.3}
From the previous section, when a bivariate normal distribution can be assumed as the latent distribution of a two-way contingency table, the power-divergence is defined as follows:
\begin{align}
I_\lambda(\{p_{ij}\};\{p_{i \cdot}p_{\cdot j}\}) &= \frac{1}{\lambda(\lambda + 1)} \left\lbrace (1-\rho^2_{(\lambda)})^{-\frac{\lambda}{2}}(1-\lambda^2 \rho^2_{(\lambda)})^{-\frac{1}{2}}-1 \right\rbrace, \quad -1 \leq \lambda \leq 1,
\label{Eq1}
\end{align}
where, $\rho_{(\lambda)}$ denotes the correlation coefficient written to distinguish the parameter of the divergence being applied.
The power divergence is also a member of the generalized divergence family, and for $-1 \leq \lambda \leq 1$ it coincides with the reverse KL, Freeman-Tukey (FT), KL, Cressie-Read (CR), and Pearson divergences at $\lambda = -1, -1/2, 0$, $2/3$, and $1$ respectively, where the cases $\lambda = -1$ and $0$ are understood in the limit as $\lambda \to -1$ and $\lambda \to 0$.
Equation (\ref{Eq1}) shows that various divergences, including several well-known ones, have a connection with the correlation coefficient. 
As special cases, we consider the relationship between the correlation coefficient and the KL and Pearson divergences obtained by setting the parameter $\lambda = 0$ and $1$. 
Solving Eq. (\ref{Eq1}) for $\rho^{2}_{(\lambda)}$ at each of these parameter values yields the following association measures for contingency tables whose latent distribution is assumed to be bivariate normal.
\begin{align*}
\rho^2_{(0)} &= 1 - \exp\left[-2I_{KL}(\{p_{ij}\}; \{p_{i\cdot} p_{\cdot j}\})\right], \\
\rho^2_{(1)} &= 1 - \left[2I_{P}(\{p_{ij}\}; \{p_{i\cdot} p_{\cdot j}\}) +1 \right]^{-1}
%\rho^2_{(0)} &= 1 - \exp\left(-2\sum^r_{i=1}\sum^c_{j=1} p_{ij}\log{\frac{p_{ij}}{p_{i\cdot} p_{\cdot j}}}\right), \\
%\rho^2_{(1)} &= 1 - \left\{\sum^r_{i=1}\sum^c_{j=1} \frac{(p_{ij}-p_{i\cdot} p_{\cdot j})^2}{p_{i\cdot} p_{\cdot j}} +1 \right\}^{-1}.
\end{align*}
where $I_{KL}(\cdot; \cdot)$ and $I_{P}(\cdot; \cdot)$ are the KL and the Pearson divergence, and are expressed as:
\begin{align*}
I_{KL}(\{p_{ij}\}; \{p_{i\cdot} p_{\cdot j}\}) &= \sum^r_{i=1}\sum^c_{j=1} p_{ij}\log{\frac{p_{ij}}{p_{i\cdot} p_{\cdot j}}}, \\
I_{P}(\{p_{ij}\}; \{p_{i\cdot} p_{\cdot j}\}) &= \frac{1}{2}\sum^r_{i=1}\sum^c_{j=1} \frac{(p_{ij}-p_{i\cdot} p_{\cdot j})^2}{p_{i\cdot} p_{\cdot j}}.
\end{align*}
For these measures, the following theorem is obtained.

\begin{theorem}\label{the1}
When $H(X)$ and $H(Y)$ are the Shannon entropy of row and column variables, respectively, the measure $\rho^2_{(0)}$ satisfies the following three properties:
\begin{itemize}
\item[1] $0 \leq \rho^2_{(0)} \leq 1-\exp\left\{-2\min\left(H(X), H(Y) \right)\right\}$.
\begin{itemize}
\item[1-1] When $r \geq c$ and $p_{i\cdot} = r^{-1}$, $0 \leq \rho^2_{(0)} \leq 1-r^{-2}$.
\item[1-2] When $r < c$ and $p_{\cdot j} = c^{-1}$, $0 \leq \rho^2_{(0)} \leq 1-c^{-2}$.
\end{itemize}
\item[2] $\rho^2_{(0)}=0$ if and only if there is a structure of independence. %$($i.e., $\{p_{ij} = p_{i \cdot}p_{\cdot j}\})$.
\item[3] $\rho^2_{(0)}$ is maximum value if and only if there is a structure of complete association.
\item[4] When $r, c \rightarrow \infty$, $0 \leq \rho^2_{(0)} < 1$
\end{itemize}
\end{theorem}

\begin{theorem}\label{the2}
The measure $\rho^2_{(1)}$ satisfies the following three properties:
\begin{itemize}
\item[1] $0 \leq \rho^2_{(1)} \leq 1-\min(r, c)^{-1}$.
\item[2] $\rho^2_{(1)}=0$ if and only if there is a structure of independence. %$($i.e., $\{p_{ij} = p_{i \cdot}p_{\cdot j}\})$.
\item[3] $\rho^2_{(1)}$ is maximum value if and only if there is a structure of complete association.%; namely, 
\item[4] When $r, c \rightarrow \infty$, $0 \leq \rho^2_{(1)} < 1$
\end{itemize}
\end{theorem}

For these two special cases, the case $\lambda = 1$ corresponds to an association measure, and the case $\lambda = 0$ corresponds to a pseudo coefficient of determination, both of which have already been proposed in the existing literature. 
In particular, $\rho^2_{(1)}$ coincides with Pearson's contingency coefficient $C$ proposed in \cite{pearson1904theory}, which linked a latent correlation coefficient with an association measure for the two-way $r \times c$ contingency table before the tetrachoric and polychoric correlation coefficients were introduced.
Furthermore, $\rho^2_{(0)}$ is consistent with the square of informational measure of correlation (\citealp{linfoot1957informational}) and, from the perspective of a pseudo coefficient of determination, coincides with Cox-Snell's $R^2$ (\citealp{cox1989analysis}) for a multinomial logit model with multi-valued categorical explanatory variable $X$ and response variable $Y$, compared with the null model with no predictors.
That is, letting $L_1$ denote the likelihood for the multinomial logit model and the likelihood $L_0$ for the null model, then
\begin{align*}
R^2_{CS} &= 1-exp\left\{-\frac{2}{n}\log\left(\frac{L_1}{L_0} \right) \right\} \\ &= 1-exp\left\{-\frac{2}{n}I_{KL}(\{\hat{p}_{ij}\}; \{\hat{p}_{i\cdot} \hat{p}_{\cdot j}\}) \right\} \\
&= \hat{\rho}^2_{(0)}
\end{align*}
where $\hat{\rho}^2_{(0)}$ is the plug-in estimator of $\rho^2_{(0)}$ obtained by replacing $\{p_{ij}\}$ with $\{\hat{p}_{ij}\}$.
The limitations of Cox-Snell's $R^2$ are also reflected in Theorem \ref{the1}, such as the fact that its maximum value is $0.75$ when response variable $Y$ is binary with $\Pr(Y=0)=\Pr(Y=1)=0.5$, so it never reaches $1$, and that its upper bound depends on the probability distribution of the response variable.
In addition, when an association measure does not reach $1$, it is common to standardize it by dividing by its upper bound. 
Applying this idea to $\hat\rho^2_{(0)}$ shows that it coincides with Nagelkerke's pseudo $R^2$ (\citealp{nagelkerke1991note}), which was proposed as an improvement to the limitations of Cox-Snell's $R^2$.

\section{Proposed divergence-based correlation measures}
In this section, based on the relationship introduced in Section 2, we present an approach for computing the proposed correlation measure $\rho_{(\lambda)}$ over the parameter range $-1 \leq \lambda \leq 1$. 
We also describe the asymptotic confidence intervals constructed using the simple method and Fisher's z method.
For simplicity, we vectorize $\{p_{ij}\}$ and $\{\hat p_{ij}\}$ as 
\begin{align*}
\bm{p} = (p_{11}, p_{12}, \dots, p_{r(c-1)}, p_{rc})^T \quad \text{and} \quad \hat{\bm{p}} = (\hat{p}_{11}, \hat{p}_{12}, \dots, \hat{p}_{r(c-1)}, \hat{p}_{rc})^T.
\end{align*} 
We then define the power divergence as functions 
\begin{align*}
D_\lambda(\bm{p}) = I_\lambda(\{p_{ij}\};\{p_{i \cdot}p_{\cdot j}\}) \quad \text{and} \quad D_\lambda(\hat{\bm{p}}) = I_\lambda(\{\hat{p}_{ij}\};\{\hat{p}_{i \cdot}\hat{p}_{\cdot j}\}) 
\end{align*}
of $\bm{p}$ and $\hat{\bm{p}}$.
Let $t_{(\lambda)} = \rho_{(\lambda)}^{2}$ and its estimator $\hat t_{(\lambda)} = \hat \rho_{(\lambda)}^{2}$.
For $0 \leq t < 1$, we define the function 
\begin{align*}
I_\lambda(t) &= \frac{1}{\lambda(\lambda + 1)} \left\{ (1-t)^{-\frac{\lambda}{2}} (1-\lambda^2 t)^{-\frac{1}{2}} - 1 \right\}.
\end{align*}
In particular, under the assumption that the latent distribution of the contingency table is bivariate normal, we assume that 
\begin{align*}
I_\lambda(t_{(\lambda)}) &= D_\lambda(\bm p) \quad \text{and} \quad I_\lambda(\hat t_{(\lambda)})= D_\lambda(\hat{\bm p}).
\end{align*}
\begin{comment}
, and define their functions of the correlation coefficient as
\begin{align*}
I_\lambda(t_{(\lambda)}) &= \frac{1}{\lambda(\lambda + 1)} \left\lbrace (1-t_{(\lambda)})^{-\frac{\lambda}{2}}(1-\lambda^2 t_{(\lambda)})^{-\frac{1}{2}}-1 \right\rbrace
\end{align*}
and 
\begin{align*}
I_\lambda(\hat{t}_{(\lambda)}) &= \frac{1}{\lambda(\lambda + 1)} \left\lbrace (1-\hat{t}_{(\lambda)})^{-\frac{\lambda}{2}}(1-\lambda^2 \hat{t}_{(\lambda)})^{-\frac{1}{2}}-1 \right\rbrace.
\end{align*}

Under the assumption that the latent distribution of the contingency table is bivariate normal, we assume that 
\begin{align*}
I_\lambda(\bm{p}) = I_\lambda(t_{(\lambda)}) 
\quad \text{and} \quad 
I_\lambda(\hat{\bm{p}}) = I_\lambda(\hat{t}_{(\lambda)})
\end{align*}
hold.
\end{comment}

%%%%%%%%%%%%%%%%%%%%%%%%%%%%%%%%%%%%%%%%%%%%%%%%%%
\subsection{Numerical computation via Newton's method}
In this subsection, we describe a numerical procedure for obtaining $\hat{t}_{(\lambda)}=\hat\rho_{(\lambda)}^2$ from the empirical power-divergence $I_\lambda(\hat{\bm p})$.
For $-1\leq\lambda\leq1$, recall that $t_{(\lambda)}$ and $\hat t_{(\lambda)}$ are implicitly defined by
\begin{align*}
I_\lambda(t_{(\lambda)}) &= D_\lambda(\bm p) \quad \text{and} \quad I_\lambda(\hat t_{(\lambda)})= D_\lambda(\hat{\bm p}).
\end{align*}
Thus, for a fixed $\lambda$ and a given $\hat D_\lambda := D_\lambda(\hat{\bm p})\geq0$,
the estimator $\hat t_{(\lambda)}$ is the unique solution of
\begin{align}
\label{eq:newton-root}
I_\lambda(t) = \hat D_\lambda, \quad 0\leq t < 1.
\end{align}

As shown in Appendix \ref{A1}, $I_\lambda(t)$ is strictly increasing on $t \in [0,1)$ for
$-1\leq\lambda\leq1$, with $I_\lambda(0)=0$ and $\lim_{t\rightarrow 1}I_\lambda(t)=\infty$.
Therefore, for each $\hat D_\lambda\geq0$ the equation \eqref{eq:newton-root} admits a unique solution $\hat t_{(\lambda)}\in[0,1)$, and we set $\hat\rho_{(\lambda)} = \sqrt{\hat t_{(\lambda)}}$.

To compute $\hat t_{(\lambda)}$ we apply Newton's method to 
\begin{align*}
f(t) = I_\lambda(t) - \hat D_\lambda.
\end{align*}
When we write the derivative of $f(t)$ as 
\begin{align*}
f'(t) = I'_\lambda(t) &= \frac{S_{\lambda}(t)}{2(\lambda+1)}\left(\frac{1}{1-t} + \frac{\lambda}{1-\lambda^2t} \right),
\end{align*}
where
\begin{align*}
S_{\lambda}(t) &= (1-t)^{-\frac{\lambda}{2}}(1-\lambda^2 t)^{-\frac{1}{2}},
\end{align*}
the Newton update is given by
\begin{equation}
\label{eq:newton-main}
t_{k+1} = t_k - \frac{f(t_k) }{f'(t_k)}, \quad k=0,1,2,\dots .
\end{equation}
A suitable initial value $t_0$ is defined as %$t_0$ is obtained from the small-$t$ expansion of $I_\lambda(t)$,
\begin{align*}
t_0 = 2\hat D_\lambda - (3\lambda^2-\lambda+2)\,\hat D_\lambda^2.  
\end{align*}
and the derivation of $t_0$ is shown in Appendix \ref{A2}.
In practice, starting from $t_0$, we iterate the Newton update (\ref{eq:newton-main}) until one of the following stopping criteria is met:
\begin{align*}
|t_{k+1}-t_k|< \delta.
\end{align*}
In our numerical experiments, we set $\delta=10^{-8}$.
Whenever the raw Newton update falls outside the interval $[0,1)$, we replace it by a projection onto $[0,1-\varepsilon]$ with a small $\varepsilon > 0$.

For $\lambda=0$ (KL divergence) and $\lambda=1$ (Pearson divergence), closed-form expressions of $\rho_{(\lambda)}$ in terms of $I_\lambda$ are available and can be used instead of the numerical solution.
Nevertheless, the Newton scheme (\ref{eq:newton-main}) provides a unified implementation for all $-1\leq\lambda\leq1$.

\begin{comment}
\begin{figure}[!t]
\begin{algorithm}[H]
\caption{Computation of the estimate $\hat\rho_{(\lambda)}$}
\begin{algorithmic}[1]
\REQUIRE $\hat D_\lambda$, divergence parameter $\lambda$
\STATE Compute initial value
  $t_0 = 2\hat D_\lambda - (3\lambda^2-\lambda+2)\hat D_\lambda^2$
  and truncate to $[0,1-\varepsilon]$
\STATE Set $k \gets 0$
\REPEAT
  \STATE Compute $f_\lambda(t_k)=I_\lambda(t_k)-\hat D_\lambda$
  and $f'_\lambda(t_k)=I'_\lambda(t_k)$
  \STATE Update
    $t_{k+1} \gets t_k - f_\lambda(t_k)/f'_\lambda(t_k)$
    and truncate to $[0,1-\varepsilon]$ if necessary
  \STATE $k \gets k+1$
\UNTIL{$|t_k-t_{k-1}|<10^{-8}$ or $|f_\lambda(t_k)|<10^{-10}$}
\STATE \textbf{output} $\hat\rho_{(\lambda)} = \sqrt{t_k}$
\end{algorithmic}
\end{algorithm}
\end{figure}
\end{comment}

%%%%%%%%%%%%%%%%%%%%%%%%%%%%%%%%%%%%%%%%%%%%%%%%%%
\subsection{Asymptotic distributions and confidence intervals}\label{Sec3.2}
In this subsection, we introduce two approaches for constructing asymptotic confidence intervals for $\rho_{(\lambda)}$ over the range $-1 \leq \lambda \leq 1$. 
The first approach constructs asymptotic confidence intervals by applying the delta method (\citealp{agresti2010analysis}) directly to the asymptotic distributions of $\rho^2_{(\lambda)}$ and $\rho_{(\lambda)}$. 
The second approach applies Fisher’s $Z$ transformation to $\rho_{(\lambda)}$ and then uses the delta method to obtain the asymptotic distribution of the transformed statistic, from which the confidence intervals are constructed. 
Following \cite{itaya2025asymptotic}, we refer to the first approach as the ``Simple method'' and the second approach as the ``Fisher’s z method''.

\subsubsection{Simple method}
The first method constructs an approximate confidence interval by applying the delta method directly to $\hat t_{(\lambda)}$ to obtain its asymptotic distribution. 
For this purpose, it is necessary to derive the asymptotic distribution of the estimator of $D_\lambda(\bm p)$ using the delta method, and Theorem \ref{thm3} provides the asymptotic distribution of $\hat t_{(\lambda)}$.
\begin{theorem}\label{thm3}
Let $\hat t_{(\lambda)}$ be the estimator for $t_{(\lambda)}$.
Then, 
\begin{align*}
\sqrt{n}\left(\hat t_{(\lambda)} - t_{(\lambda)}\right) \xrightarrow{d} N\left(0, \sigma^2_{t} \right),
\end{align*}
where, 
\begin{align*}
\sigma^2_{t} = \frac{\sigma^2_{D_{\lambda}}}{\left(I'_\lambda(t_{(\lambda)})\right)^2}
\end{align*}
and 
\begin{align*}
\sigma^2_{D_{\lambda}} &= \nabla D_\lambda(\bm{p})^T\;(diag(\bm{p})-\bm{p}\bm{p}^T) \; \nabla D_\lambda(\bm{p})
\end{align*}
\end{theorem}
For the proof of Theorem \ref{thm3} and the notation for $\sigma^2_{D_{\lambda}}$, see the Appendix \ref{A3}.
Thus, a $100(1-\alpha)\%$ approximate confidence interval for $t_{(\lambda)} = \rho^2_{(\lambda)}$ is given by
\begin{align*}
\left[\,\hat t_{(\lambda)} - z_{\alpha/2}\times\sqrt{\frac{\sigma^2_{t}}{n}}, \quad \hat t_{(\lambda)} + z_{\alpha/2}\times\sqrt{\frac{\sigma^2_{t}}{n}}\,\right]
\end{align*}
where $z_{\alpha/2}$ is the upper $\alpha/2$-quantile of the standard normal distribution.
However, in practical data analysis, we require an interval for $\rho_{(\lambda)} = \sqrt{t_{(\lambda)}}$, so we provide the following corollary.
\begin{corollary}
Let $\hat{\rho}_{(\lambda)}=\sqrt{\hat t_{(\lambda)}}$ be the estimator for the measure $\rho_{(\lambda)}=\sqrt{t_{(\lambda)}}$.
Then, 
\begin{align*}
\sqrt{n}\left(\hat{\rho}_{(\lambda)}-\rho_{(\lambda)}\right) \xrightarrow{d} N\left(0, \sigma^2_{\rho} \right),
\end{align*}
where, 
\begin{align*}
\sigma^2_{\rho} = \frac{\sigma^2_{t}}{4\rho^2_{(\lambda)}} = \frac{\sigma^2_{D_{\lambda}}}{4\rho^2_{(\lambda)}\left(I'_\lambda(t_{(\lambda)})\right)^2}.
\end{align*}
\end{corollary}
Accordingly, a $100(1-\alpha)\%$ approximate confidence interval for $\rho_{(\lambda)}$ is given by
\begin{align*}
\left[\,\hat \rho_{(\lambda)} - z_{\alpha/2}\times\sqrt{\frac{\sigma^2_{\rho}}{n}}, \quad \hat \rho_{(\lambda)} + z_{\alpha/2}\times\sqrt{\frac{\sigma^2_{\rho}}{n}}\,\right],
\end{align*}
Notably, when $\lambda = 1$, the asymptotic distribution obtained via the delta method coincides with that for Pearson’s contingency coefficient $C$. 
This shows that the proposed results provide asymptotic distributions and approximate confidence intervals that generalize those of Pearson’s contingency coefficient $C$.

\subsubsection{Fisher's $z$ method}
The second method constructs an approximate confidence interval by applying Fisher’s $z$ transformation to $\rho^2_{(\lambda)}$ and then introducing the asymptotic distribution via the delta method. 
%Fisher’s $z$ transformation is widely used in the asymptotic theory of Pearson’s correlation coefficient when constructing confidence intervals, and it is known to have several desirable properties, including variance stabilization.
Fisher's z transformation is widely used in the construction of confidence intervals for the Pearson correlation coefficient, and we apply it here in an analogous fashion, though as shown below, variance stabilization does not carry over to the present setting.
%Motivated by this, we apply the same transformation in our setting and define it as:
\begin{align*}
z = \text{artanh}(\rho_{(\lambda)}) =\frac{1}{2}\log\left(\frac{1+\rho_{(\lambda)}}{1-\rho_{(\lambda)}}\right).
\end{align*}
It is well known that the asymptotic variance of Pearson’s correlation coefficient under Fisher’s $z$ transformation is approximated by $1/(n-3)$. 
However, in the present setting, the transformation does not achieve variance stabilization; the asymptotic variance retains dependence on $\rho_{(\lambda)}$. 
Therefore, by applying the delta method, we obtain the asymptotic distribution given in Theorem \ref{thm4}.
\begin{theorem}\label{thm4}
Let $\hat{z}$ be the estimator for the z-transformation $z$.
Then, 
\begin{align*}
\sqrt{n}\left(\hat{z}-z\right) \xrightarrow{d} N\left(0, \sigma^2_{z}\right),
\end{align*}
where
\begin{align*}
\sigma^2_{z} = \frac{\sigma^2_{\rho}}{\left(1-\rho^2_{(\lambda)}\right)^2} = \frac{\sigma^2_{t}}{4\rho^2_{(\lambda)}\left(1-\rho^2_{(\lambda)}\right)^2} = \frac{\sigma^2_{D_{\lambda}}}{4\rho^2_{(\lambda)}\left(1-\rho^2_{(\lambda)}\right)^2\left(I'_\lambda(t_{(\lambda)})\right)^2} 
\end{align*}
\end{theorem}
Thus, the $100(1-\alpha)\%$ approximate confidence interval for $z$ is given by \begin{align*}
\left[\,\hat z - z_{\alpha/2}\times\sqrt{\frac{\sigma^2_{z}}{n}}, \quad \hat z + z_{\alpha/2}\times\sqrt{\frac{\sigma^2_{z}}{n}}\,\right],
\end{align*}
and the $100(1-\alpha)\%$ approximate confidence interval for $\rho_{(\lambda)}$ obtained by applying the inverse transformation of $z$ is
\begin{align*}
\left[\,\frac{e^{2L}-1}{e^{2L}+1}, \quad \frac{e^{2U}-1}{e^{2U}+1}\,\right],
\end{align*}
where $L$ and $U$ are the lower and upper bounds, respectively, and are defined as
\begin{align*}
L = \hat z - z_{\alpha/2}\times\sqrt{\frac{\sigma^2_{z}}{n}} \quad \text{and} \quad U = \hat z + z_{\alpha/2}\times\sqrt{\frac{\sigma^2_{z}}{n}}
\end{align*}

Moreover, for the asymptotic variances proposed in these theorems and the corollary, it is clear from the fact that $\rho_{(\lambda)} \in [0,1)$ that
\begin{align*}
\sigma^2_{z} > \sigma^2_{\rho} > \sigma^2_{t}
\end{align*}
holds.

%%%%%%%%%%%%%%%%%%%%%%%%%%%%%%%%%%%%%%%%%%%%%%%%%%
\subsection{Detectable correlation thresholds based on test statistics}\label{Sec3.3}
%To ensure the stability of proposed statistical indices, including correlation coefficients and association measures, confidence intervals have been developed for each index.
While confidence intervals indicate whether an association is statistically significant, they do not convey whether its magnitude is practically meaningful. 
In this subsection, we therefore introduce detectable correlation thresholds that translate the chi-squared critical value of the independence test into a minimum value of $\rho_{(\lambda)}$ that can be regarded as practically meaningful.
In practical data analysis, constructing a confidence interval and confirming that the interval does not include zero allows us to statistically evaluate that a relationship between variables does exist. 
However, this evaluation only indicates the existence of a relationship and does not assess whether the degree of association is practically meaningful.
Therefore, in this subsection, recalling that the test statistic for independence is constructed based on a divergence (that is, $2n\hat D_{\lambda}$), we introduce a threshold that reflects the degree of practically meaningful association under significance level $\alpha$ and sample size $n$, by setting the initial value $t_0$ to
\begin{align*}
t_0 = 2 D_{\lambda}(\alpha,n) - (3\lambda^2-\lambda+2)\,D^2_{\lambda}(\alpha,n),
\end{align*}
where
\begin{align*}
D_{\lambda}(\alpha,n) = \frac{\chi^2_{\alpha}}{2n}
\end{align*}
and $\chi^2_{\alpha}$ is the upper $\alpha$-quantile of the chi-squared distribution with degree of freedom $(r-1)(c-1)$.

Tables \ref{tab:rho_FT} and \ref{tab:rho_P} report thresholds for several degrees of freedom under significance levels $\alpha = 5\%$ and $1\%$, with sample sizes $n = 1000$, $3000$, and $5000$. 
Table \ref{tab:rho_FT}  summarizes the case $\lambda = -1/2$ (related to the FT statistic), and Table \ref{tab:rho_P} summarizes the case $\lambda = 1$ (related to Pearson’s chi-squared test statistic). 
Thresholds for other values of $\lambda$ are provided in Appendix \ref{A4}.
For example, in the case of degrees of freedom equal to 4 (that is, a $3 \times 3$ contingency table) with $n = 1000$, the tables suggest that in addition to checking that the confidence intervals for $\lambda = -1/2$ and $\lambda = 1$ do not contain zero, one should also verify that they do not include $0.0970$ and $0.0969$, respectively, in order to conclude that a statistically meaningful association is present.
That said, when association measures are used in practical data analysis, it is often the case that a test of independence has already been conducted and independence has been rejected. 
If such a check has been made, it is reasonable to assume that the thresholds have been exceeded. 
However, when the analysis relies solely on association measures, it is advisable to verify the results against these thresholds.

\begin{table}[ht]
\centering
\caption{Thresholds of $\rho_{(-1/2)}$ based on the FT statistic} 
\label{tab:rho_FT}
\begin{tabular}{c|cccccc}
\hline  
& \multicolumn{2}{c}{$n=1000$} & \multicolumn{2}{c}{$n=3000$} & \multicolumn{2}{c}{$n=5000$} \\
df & $\alpha=0.05$ & $\alpha=0.01$ & $\alpha=0.05$ & $\alpha=0.01$ & $\alpha=0.05$ & $\alpha=0.01$ \\ \hline
 1 & 0.0619 & 0.0812 & 0.0358 & 0.0470 & 0.0277 & 0.0364 \\  
 2 & 0.0772 & 0.0956 & 0.0447 & 0.0553 & 0.0346 & 0.0429 \\  
 3 & 0.0881 & 0.1060 & 0.0510 & 0.0614 & 0.0395 & 0.0476 \\  
 4 & 0.0970 & 0.1146 & 0.0562 & 0.0664 & 0.0435 & 0.0515 \\ 
 5 & 0.1047 & 0.1221 & 0.0607 & 0.0708 & 0.0470 & 0.0549 \\ 
 6 & 0.1116 & 0.1288 & 0.0647 & 0.0747 & 0.0501 & 0.0579 \\ 
 7 & 0.1179 & 0.1349 & 0.0683 & 0.0783 & 0.0530 & 0.0607 \\ 
 8 & 0.1238 & 0.1406 & 0.0717 & 0.0816 & 0.0556 & 0.0633 \\ 
 9 & 0.1292 & 0.1459 & 0.0749 & 0.0847 & 0.0581 & 0.0657 \\ 
10 & 0.1343 & 0.1509 & 0.0779 & 0.0877 & 0.0604 & 0.0680 \\ 
15 & 0.1565 & 0.1727 & 0.0910 & 0.1005 & 0.0706 & 0.0780 \\ 
20 & 0.1750 & 0.1909 & 0.1019 & 0.1113 & 0.0791 & 0.0864 \\ 
25 & 0.1912 & 0.2068 & 0.1115 & 0.1208 & 0.0865 & 0.0938 \\  \hline  
\end{tabular}
\end{table}

\begin{table}[ht]
\centering
\caption{Thresholds of $\rho_{(1)}$ based on Pearson’s chi-squared test statistic} 
\label{tab:rho_P}
\begin{tabular}{c|cccccc}
\hline  
& \multicolumn{2}{c}{$n=1000$} & \multicolumn{2}{c}{$n=3000$} & \multicolumn{2}{c}{$n=5000$} \\
df & $\alpha=0.05$ & $\alpha=0.01$ & $\alpha=0.05$ & $\alpha=0.01$ & $\alpha=0.05$ & $\alpha=0.01$ \\ \hline
 1 & 0.0619 & 0.0812 & 0.0358 & 0.0470 & 0.0277 & 0.0364 \\  
 2 & 0.0772 & 0.0955 & 0.0446 & 0.0553 & 0.0346 & 0.0429 \\  
 3 & 0.0881 & 0.1059 & 0.0510 & 0.0614 & 0.0395 & 0.0476 \\  
 4 & 0.0969 & 0.1145 & 0.0561 & 0.0664 & 0.0435 & 0.0515 \\ 
 5 & 0.1046 & 0.1219 & 0.0606 & 0.0707 & 0.0470 & 0.0548 \\ 
 6 & 0.1115 & 0.1286 & 0.0647 & 0.0747 & 0.0501 & 0.0579 \\ 
 7 & 0.1178 & 0.1347 & 0.0683 & 0.0782 & 0.0530 & 0.0607 \\ 
 8 & 0.1236 & 0.1403 & 0.0717 & 0.0816 & 0.0556 & 0.0633 \\ 
 9 & 0.1290 & 0.1456 & 0.0749 & 0.0847 & 0.0581 & 0.0657 \\ 
10 & 0.1341 & 0.1506 & 0.0779 & 0.0876 & 0.0604 & 0.0680 \\ 
15 & 0.1562 & 0.1723 & 0.0909 & 0.1004 & 0.0705 & 0.0780 \\ 
20 & 0.1745 & 0.1903 & 0.1018 & 0.1112 & 0.0790 & 0.0864 \\ 
25 & 0.1905 & 0.2060 & 0.1113 & 0.1206 & 0.0865 & 0.0937 \\  \hline  
\end{tabular}
\end{table}

\section{Simulation study}
In this section, we examine how the association measures proposed in this study behave for different parameter values and evaluate the performance of their confidence intervals. 
We also discuss, based on the results of simulation experiments, which parameter values should be used in practical data analysis when measuring association in two-way contingency tables whose latent distribution can be assumed to be a bivariate normal distribution.
In the simulations, we adopt $\lambda = -1/2, 0, 2/3,$ and $1$ as parameter values, and do not include $\lambda = -1$. 
This is because $D_\lambda(\bm p)$ may be undefined when zero cells are present, a situation that arises frequently under strong association at $\lambda = -1$.
For comparison with the proposed association measures, we use Cram\'er’s coefficient $V^2$ and the total uncertainty coefficient $U_{Total}$, both of which are symmetric association measures constructed from Pearson divergence and KL divergence, respectively.

%%%%%%%%%%%%%%%%%%%%%%%%%%%%%%%%%%%%%%%%%%%
\subsection{Performances of the proposed measures}\label{Sec4.1}
In this subsection, assuming a bivariate standard normal distribution with correlation coefficient $\rho$ as the latent distribution, we investigate how closely the proposed measures and the comparison measures approximate the true correlation $\rho$. 
This assessment is conducted across several sample sizes and numbers of categories.
For constructing the tables, we generate $r \times r$ probability tables ($r = 2, \dots, 15$) by partitioning the bivariate standard normal distributions with $\rho = 0.2, 0.5,$ and $0.8$ using thresholds $z_{1/r}, \dots, z_{(r-1)/r}$ so that the row and column marginal probabilities are equal.
Based on each probability table, we then generate $r \times r$ contingency tables using multinomial sampling with sample sizes $n = 3000, 5000,$ and $10000$, producing $1000$ such tables for each setting.
The association measures are applied both to the true probability tables and to the sampled contingency tables, allowing us to examine their behavior under the true structure and under sampled tables. 
For the contingency tables, we use the mean of the $1000$ computed association measure values in each setting for evaluation.

\begin{figure}[htpb]
\centering
\includegraphics[width=1\linewidth]{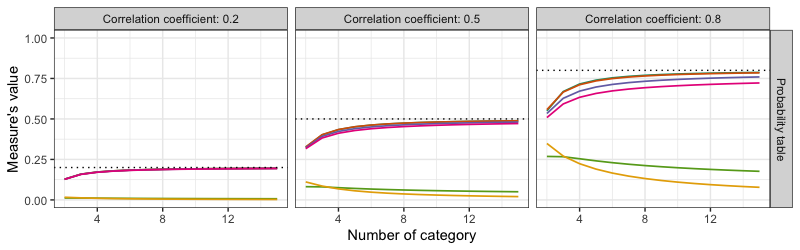}
\label{fig:PTable}
\centering
\includegraphics[width=1\linewidth]{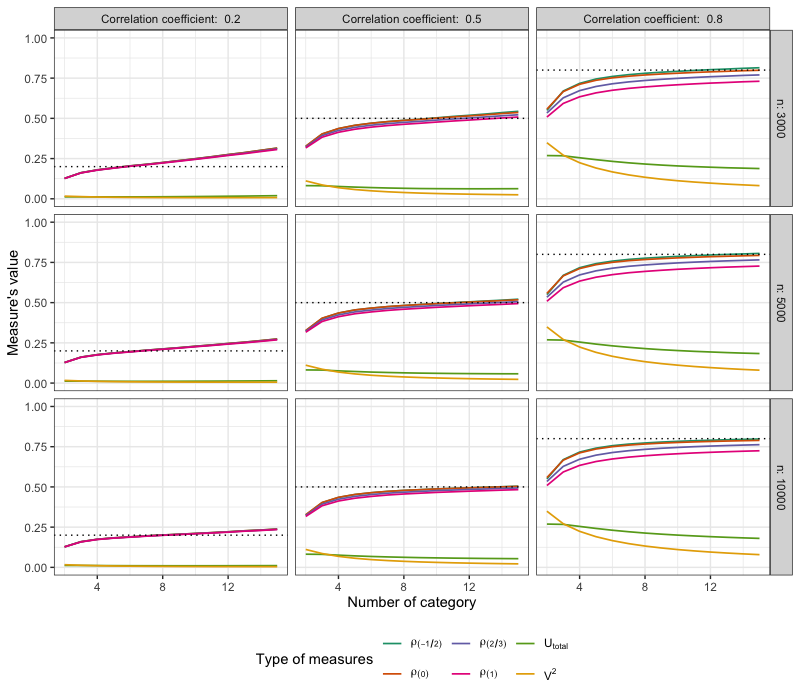}
\caption{Performances of proposed measures $\rho_{(-1/2)}$, $\rho_{(0)}$, $\rho_{(2/3)}$, and $\rho_{(1)}$, and comparison measures Cram\'er's coefficient $V^2$ and Total uncertainty measure $U_{total}$}
\label{fig:CTable}
\end{figure}

Figure \ref{fig:CTable} presents the results of the investigation. 
The upper panels show the behavior of the association measures under the true probability structure, while the lower panels show their behavior for sampled contingency tables under each setting.
From the upper panels, we see that the proposed measures approach the black dashed line representing the true correlation coefficient $\rho$ as the number of categories increases. 
This confirms the approximation behavior established in Section \ref{Sec2.2}.
However, when the underlying association is strong, such as $\rho = 0.8$, the discrepancy from $\rho$ grows slightly as the parameter value increases.
In contrast, Cram\'er's coefficient $V^2$ and the total uncertainty coefficient $U_{Total}$ do not reflect $\rho$ well. 
In addition, the existing measures $V^2$ and $U_{Total}$ show little difference between $\rho=0.2$ and $0.5$, suggesting that they are unable to reliably distinguish weak from moderate associations.
Turning to the lower panels, which show the behavior for sampled tables, we observe that as both the sample size $n$ and the number of categories $r$ increase, the measures exhibit behavior similar to that in the upper panels.
However, when the association is weak, such as $\rho = 0.2$, and the number of categories is large relative to the sample size (for example, $n = 3000$), the proposed measures tend to overestimate the association. 
In such settings, even a weak association leads to counts concentrated along the diagonal cells rather than spreading to off-diagonal cells, which inflates the computed association measures above the true value $\rho = 0.2$.

In summary, compared with existing divergence-based measures, the proposed association measures returned values close to the true correlation coefficient $\rho$, and successfully distinguish between different strengths of association, such as $\rho = 0.2$ and $0.5$. 
When the true association is strong, $\lambda = -1/2$ and $0$ tend to produce values closest to $\rho$, whereas $\lambda = 1$ yields the largest deviation among $\rho_{(\lambda)}$. 
Furthermore, when the number of categories is large relative to the sample size, it is necessary to be careful because the measures may overestimate the association even though the true association is weak.

%%%%%%%%%%%%%%%%%%%%%%%%%%%%%%%%%%%%%%%%%%%
\subsection{Coverage probability of asymptotic confidence intervals}\label{Sec4.2}
In this subsection, we constructed $95\%$ approximate confidence intervals $100,000$ times using the two methods proposed in Section \ref{Sec3.2}, the ``Simple method'' and the ``Fisher’s $z$ method'', and examined the coverage probabilities for each parameter value. 
For generating the $100,000$ sets of approximate confidence intervals for each parameter value, we adopted the same settings for $\rho$ and $n$ as in Section \ref{Sec4.1} and generated $4\times 4$, $6\times 6$, and $8\times 8$ contingency tables $100,000$ times following the same procedure.

\begin{table}[ht]
\centering
\caption{Coverage probabilities of the 95\% confidence intervals under the Simple method and Fisher's $z$ method} 
\label{tab:cp}
\begin{tabular}{ccc|cccc|cccc}
\hline  
 & & & \multicolumn{4}{c}{Simple method} & \multicolumn{4}{c}{Fisher’s $z$ method}  \\
$r$ & $\rho$ & $n$ & $\rho_{(-1/2)}$ & $\rho_{(0)}$ & $\rho_{(2/3)}$ & $\rho_{(1)}$ & $\rho_{(-1/2)}$ & $\rho_{(0)}$ & $\rho_{(2/3)}$ & $\rho_{(1)}$ \\ \hline
4 & 0.2
%     &  1000 & 0.89802 & \textbf{0.89828} & 0.89741 & 0.89675 
%             & 0.90245 & \textbf{0.90276} & 0.90187 & 0.90077 \\  
     &  3000 & \textbf{0.93280} & 0.93261 & 0.93236 & 0.93207 
             & 0.93408 & \textbf{0.93419} & 0.93358 & 0.93321 \\  
  &  &  5000 & \textbf{0.93882} & 0.93877 & 0.93855 & 0.93828 
             & \textbf{0.93976} & 0.93972 & 0.93942 & 0.93913\\  
  &  & 10000 & 0.94491 & 0.94495 & 0.94520 & \textbf{0.94522} 
             & 0.94553 & 0.94540 & \textbf{0.94562} & 0.94551 \\ 
  & 0.5
%     &  1000 & 0.93434 & 0.93485 & 0.93631 & \textbf{0.93660} 
%             & 0.93848 & 0.93880 & 0.93957 & \textbf{0.93971} \\  
     &  3000 & 0.94562 & 0.94573 & \textbf{0.94623} & 0.94602 
             & 0.94694 & 0.94723 & \textbf{0.94729} & 0.94721 \\  
  &  &  5000 & \textbf{0.94791} & 0.94745 & 0.94723 & 0.94733 
             & \textbf{0.94886} & 0.94805 & 0.94797 & 0.94808 \\  
  &  & 10000 & 0.94901 & 0.94934 & \textbf{0.94965} & 0.94949 
             & 0.94932 & 0.94981 & \textbf{0.95003} & 0.94977 \\ 
  & 0.8
%     &  1000 & 0.83804 & 0.93427 & 0.94449 & \textbf{0.94631} 
%             & 0.84821 & 0.93830 & 0.94738 & \textbf{0.94780} \\  
     &  3000 & 0.93225 & 0.94529 & 0.94889 & \textbf{0.94918} 
             & 0.93491 & 0.94703 & \textbf{0.94979} & \textbf{0.94979} \\  
  &  &  5000 & 0.94206 & 0.94752 & 0.94900 & \textbf{0.94902} 
             & 0.94358 & 0.94814 & 0.94925 & \textbf{0.94952} \\  
  &  & 10000 & 0.94546 & 0.94745 & \textbf{0.94859} & 0.94855 
             & 0.94603 & 0.94793 & \textbf{0.94903} & 0.94861 \\  
\hline  

  6 & 0.2
%     &  1000 & 0.54486 & 0.54542 & 0.55071 & \textbf{0.55378} 
%             & 0.55638 & 0.55699 & 0.56192 & \textbf{0.56479} \\  
     &  3000 & 0.80136 & 0.80029 & 0.80234 & \textbf{0.80436} 
             & 0.80514 & 0.80410 & 0.80626 & \textbf{0.80794} \\  
  &  &  5000 & 0.85939 & 0.85916 & 0.85987 & \textbf{0.86110} 
             & 0.86186 & 0.86140 & 0.86215 & \textbf{0.86345} \\  
  &  & 10000 & 0.90268 & 0.90231 & 0.90268 & \textbf{0.90325} 
             & 0.90392 & 0.90355 & 0.90379 & \textbf{0.90463} \\ 
  & 0.5
%     &  1000 & 0.83782 & 0.85169 & 0.86712 & \textbf{0.87452} 
%             & 0.85077 & 0.86313 & 0.87743 & \textbf{0.88338} \\  
     &  3000 & 0.91622 & 0.91872 & 0.92370 & \textbf{0.92568} 
             & 0.92046 & 0.92289 & 0.92684 & \textbf{0.92827} \\  
  &  &  5000 & 0.92926 & 0.93005 & 0.93345 & \textbf{0.93523} 
             & 0.93149 & 0.93261 & 0.93530 & \textbf{0.93672} \\  
  &  & 10000 & 0.94272 & 0.94302 & 0.94309 & \textbf{0.94337} 
             & 0.94394 & 0.94421 & 0.94391 & \textbf{0.94425} \\ 
  & 0.8
%     &  1000 & 0.65524 & 0.88959 & 0.92306 & \textbf{0.92814} 
%             & 0.67715 & 0.90161 & 0.92969 & \textbf{0.93284} \\  
     &  3000 & 0.82478 & 0.93013 & 0.94140 & \textbf{0.94318} 
             & 0.83427 & 0.93410 & 0.94385 & \textbf{0.94450} \\  
  &  &  5000 & 0.85716 & 0.93984 & 0.94593 & \textbf{0.94658} 
             & 0.86441 & 0.94171 & 0.94707 & \textbf{0.94788} \\  
  &  & 10000 & 0.87950 & 0.94324 & 0.94650 & \textbf{0.94777} 
             & 0.88307 & 0.94461 & 0.94735 & \textbf{0.94824} \\  
\hline  

  8 & 0.2
%     &  1000 & 0.05586 & 0.05621 & 0.05989 & \textbf{0.06307} 
%             & 0.05948 & 0.06019 & 0.06390 & \textbf{0.06712} \\  
     &  3000 & 0.42124 & 0.41906 & 0.42681 & \textbf{0.43376} 
             & 0.42739 & 0.42520 & 0.43284 & \textbf{0.44016} \\  
  &  &  5000 & 0.60244 & 0.60044 & 0.60623 & \textbf{0.61253} 
             & 0.60686 & 0.60463 & 0.61062 & \textbf{0.61693} \\  
  &  & 10000 & 0.76919 & 0.76743 & 0.77154 & \textbf{0.77527} 
             & 0.77143 & 0.76974 & 0.77384 & \textbf{0.77741} \\ 
  & 0.5
%     &  1000 & 0.52137 & 0.60096 & 0.66401 & \textbf{0.69512} 
%             & 0.54114 & 0.62227 & 0.68320 & \textbf{0.71246} \\  
     &  3000 & 0.82656 & 0.83586 & 0.85535 & \textbf{0.86686} 
             & 0.83428 & 0.84378 & 0.86188 & \textbf{0.87252} \\  
  &  &  5000 & 0.87799 & 0.88134 & 0.89281 & \textbf{0.89946} 
             & 0.88251 & 0.88579 & 0.89668 & \textbf{0.90319} \\  
  &  & 10000 & 0.91586 & 0.91650 & 0.92239 & \textbf{0.92538} 
             & 0.91816 & 0.91866 & 0.92425 & \textbf{0.92706} \\ 
  & 0.8
%     &  1000 & 0.34833 & 0.77737 & 0.86544 & \textbf{0.88368} 
%             & 0.37105 & 0.79858 & 0.87900 & \textbf{0.89297} \\  
     &  3000 & 0.62712 & 0.89156 & 0.92176 & \textbf{0.92761} 
             & 0.64229 & 0.89900 & 0.92617 & \textbf{0.93086} \\  
  &  &  5000 & 0.71772 & 0.91554 & 0.93340 & \textbf{0.93704} 
             & 0.72867 & 0.91947 & 0.93625 & \textbf{0.93891} \\  
  &  & 10000 & 0.80652 & 0.93394 & 0.94295 & \textbf{0.94462} 
             & 0.81301 & 0.93630 & 0.94425 & \textbf{0.94550} \\  
\hline  
\end{tabular}
\end{table}

Table \ref{tab:cp} reports the coverage probabilities for each parameter under every setting. 
For each of the two methods, the Simple method and Fisher’s $z$ method, the parameter value that attains the highest coverage is highlighted in bold.
For $r=4$, the results show that across all settings, regardless of the value of $\lambda$, the deviation from $0.95$ remains within roughly $0.01$, indicating good agreement.
Fisher’s $z$ method tends to give slightly higher coverage than the Simple method, but the differences are minor, suggesting that the two methods perform almost equivalently in practice.
However, as the number of categories increases to $r=6$ and $8$, the coverage probabilities deviate substantially from $0.95$, particularly when $\rho$ is small. 
For example, when $\rho = 0.2$ in the $8\times 8$ table, even with a relatively large sample size of $n=10,000$, the coverage probability is only about $0.77$.
Even for a moderate association, such as $\rho = 0.5$, coverage remains around $0.91$ to $0.93$, slightly below $0.95$. 
These findings show that as the table becomes sparser relative to the assumed true correlation, the performance of both approximate confidence interval methods deteriorates.
The parameter $\lambda$ also influences coverage. 
For instance, in several settings such as $r=6$ with $\rho = 0.8$, the coverage for $\rho_{(-1/2)}$ is noticeably poorer than for other parameter values. 
In contrast, when $r$ is large and association is moderate to strong, $\lambda = 1$ consistently yields the highest coverage, followed by $\lambda = 2/3$.

Overall, both methods exhibit good coverage when the sample size is sufficiently large relative to the number of categories. 
In sparse settings, however, where the sample size is small relative to the number of categories, substantial undercoverage can occur.
This suggests that the normal approximations underlying the interval construction become unreliable under sparsity, and caution is required when applying these methods in practical analysis.

\subsection{Polychoric correlation vs proposed measure in probability tables}\label{Sec4.3}
As with our proposed method, a well-known approach for estimating association in two-way contingency tables by assuming an underlying correlation coefficient is the polychoric correlation coefficient.
In this subsection, we demonstrate the advantages of the proposed method over the polychoric correlation coefficient, focusing on computational time and behavior in cases where the true correlation coefficient $\rho$ is close to one.
In this experiment, we adopt only the probability table generation procedure described in Section \ref{Sec4.1} and compare the performance of these association measures using contingency tables derived under the assumption of a bivariate normal distribution.

%%%%%%%%%%%%%%%%%%%%%%
\subsubsection{Runtime comparison across category sizes}\label{Sec4.3.1}
Following the same procedure as in Section 4.1, we constructed probability tables with equal row and column marginal probabilities. 
Specifically, for the bivariate standard normal distributions with $\rho = 0.2, 0.5,$ and $0.8$, we formed $r \times r$ probability tables ($r = 10, 15, 25,$ and $50$) based on the cumulative probabilities of the resulting categories.
For each of these settings, we examined both the computational time and the estimated values of the polychoric correlation coefficient and the proposed measures with $\lambda = -1/2, 0, 2/3,$ and $1$.
The polychoric correlation coefficient $\rho_{\text{poly(ML)}}$ and $\rho_{\text{poly}}$ were computed using the R function polychor() with the option ML = TRUE and FALSE, respectively, corresponding to the maximum likelihood estimation and the two-step estimation.

\begin{table}[ht]
\centering
\caption{(a) Computational time comparison and (b) measures' value of the polychoric correlation and the proposed measures as changing the number of categories} 
\label{tab:time_values}
\begin{tabular}{cc|cccccc}
\\
 \multicolumn{7}{c}{(a) Computational time (sec) of each setting}  \\ \hline  
$\rho$ & $r$ & $\rho_{\text{poly(ML)}}$ & $\rho_{\text{poly}}$ & $\rho_{(-1/2)}$ & $\rho_{(0)}$ & $\rho_{(2/3)}$ & $\rho_{(1)}$  \\ \hline
0.2 
  & 10 & 2.18330 & 2.18330 & 0.00030 & 0.00009 & 0.00015 & 0.00014 \\  
  & 15 & 5.35995 & 2.18330 & 0.00027 & 0.00010 & 0.00016 & 0.00015\\  
  & 25 & 12.55506 & 2.18330 & 0.00029 & 0.00011 & 0.00017 & 0.00016 \\ 
  & 50 & 32.31631 & 2.18330 & 0.00037 & 0.00016 & 0.00023 & 0.00023 \\ \hline
0.5 
  & 10 & 2.03311 & 2.18330 & 0.00026 & 0.00010 & 0.00016 & 0.00014 \\  
  & 15 & 5.28162 & 2.18330 & 0.00028 & 0.00010 & 0.00016 & 0.00015 \\ 
  & 25 & 12.51459 & 2.18330 & 0.00028 & 0.00012 & 0.00017 & 0.00016 \\ 
  & 50 & 32.54745 & 2.18330 & 0.00036 & 0.00017 & 0.00023 & 0.00023 \\ \hline
0.8 
  & 10 & 2.04339 & 2.18330 & 0.00026 & 0.00010 & 0.00015 & 0.00014 \\  
  & 15 & 5.29712 & 2.18330 & 0.00028 & 0.00012 & 0.00017 & 0.00015 \\  
  & 25 & 13.20488 & 2.18330 & 0.00033 & 0.00012 & 0.00018 & 0.00017 \\ 
  & 50 & 33.24196 & 2.18330 & 0.00071 & 0.00023 & 0.00044 & 0.00042 \\ \hline  
\end{tabular}

\begin{tabular}{cc|cccccc}
\\
 \multicolumn{7}{c}{(b) Measures' value of each setting}  \\ \hline  
$\rho$ & $r$ & $\rho_{\text{poly(ML)}}$ & $\rho_{\text{poly}}$ & $\rho_{(-1/2)}$ & $\rho_{(0)}$ & $\rho_{(2/3)}$ & $\rho_{(1)}$  \\ \hline
0.2 
  & 10 & 0.20001 & 0.20001 & 0.19183 & 0.19187 & 0.19119 & 0.19051 \\  
  & 15 & 0.20000 & 0.20000 & 0.19517 & 0.19520 & 0.19465 & 0.19411 \\  
  & 25 & 0.20000 & 0.20000 & 0.19748 & 0.19750 & 0.19711 & 0.19671 \\ 
  & 50 & 0.20000 & 0.20000 & 0.19894 & 0.19895 & 0.19872 & 0.19848 \\ \hline
0.5 
  & 10 & 0.50001 & 0.50001 & 0.48070 & 0.48070 & 0.47104 & 0.46080 \\  
  & 15 & 0.50002 & 0.50002 & 0.48861 & 0.48856 & 0.48062 & 0.47159 \\ 
  & 25 & 0.50002 & 0.50002 & 0.49406 & 0.49401 & 0.48808 & 0.48065 \\ 
  & 50 & 0.50002 & 0.50002 & 0.49750 & 0.49746 & 0.49366 & 0.48820 \\ \hline
0.8 
  & 10 & 0.80000 & 0.80000 & 0.77771 & 0.77425 & 0.74410 & 0.70509 \\  
  & 15 & 0.80001 & 0.80001 & 0.78734 & 0.78472 & 0.75937 & 0.72219 \\  
  & 25 & 0.80001 & 0.80001 & 0.79372 & 0.79194 & 0.77200 & 0.73773 \\ 
  & 50 & 0.80001 & 0.80001 & 0.79753 & 0.79654 & 0.78245 & 0.75242 \\ \hline  
\end{tabular}
\end{table}

Table \ref{tab:time_values}(a) summarizes the computational times (sec), while Table \ref{tab:time_values}(b) reports the values of each association measure for the probability tables under the respective settings.
First, from the results in Table \ref{tab:time_values}(a), we observe that, regardless of the value of the correlation coefficient $\rho$, the computational time increases with the number of categories $r$ for both measures. 
However, the proposed measure can be computed substantially faster than the polychoric correlation coefficient. 
For instance, when $r = 10$, the polychoric correlation coefficient requires approximately $2.18$ seconds, whereas the proposed measure takes at most about $0.00030$ seconds.
Moreover, for $r = 25$ and $50$, the difference in computational time becomes even more pronounced. 
In particular, compared with the computation of $\rho_{\mathrm{poly(ML)}}$, the proposed method enables an extremely fast evaluation of the strength of association.
Next, we turn to Table \ref{tab:time_values}(b), which summarizes the values of each association measure computed from the probability tables under the various settings. 
From these results, we see that the polychoric correlation coefficient recovers the true correlation coefficient $\rho$, whereas the proposed measure tends to slightly underestimate $\rho$. 
Nevertheless, the degree of underestimation is modest: depending on the choice of the parameter $\lambda$, it remains within a range of approximately $0.01$ to $0.03$, which can be regarded as acceptable for practical purposes.
Taken together, these findings suggest that the proposed method is particularly advantageous in exploratory analyses where a large number of contingency tables with many categories are obtained, such as in questionnaire-based surveys. 
In such settings, where fast and reliable screening of strong associations between pairs of variables is required, the proposed measure provides a computationally efficient and practically useful alternative.

%%%%%%%%%%%%%%%%%%%%%%
\subsubsection{Numerical instability near the boundary correlations}%\label{Sec4.3.2}
Next, we investigated the performance of the polychoric correlation coefficient and the proposed measures in cases where the true correlation coefficient $\rho$ is close to one.
As in Section \ref{Sec4.1}, probability tables were constructed so that the row and column marginal probabilities were equal. 
Specifically, $r \times r$ probability tables were generated from bivariate standard normal distributions with $\rho = 0.9, 0.95,$ and $0.99$, where $r$ takes the values $5, 10, 15$, and up to $100$.
For each of these settings, we evaluated the performance of the polychoric correlation coefficient and the proposed measure with $\lambda = -1/2, 0, 2/3,$ and $1$. 
In addition, following Section \ref{Sec4.1}, we also examined Cram\'er’s coefficient $V^2$ and total uncertainty measure $U_{total}$.

\begin{figure}[htpb]
\label{fig:PTable2}
\centering
\includegraphics[width=1\linewidth]{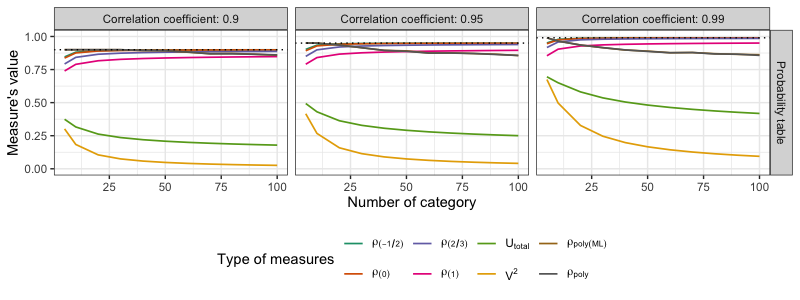}
\caption{Performances of proposed measures $\rho_{(-1/2)}$, $\rho_{(0)}$, $\rho_{(2/3)}$, and $\rho_{(1)}$, and comparison measures Cram\'er's coefficient $V^2$, Total uncertainty measure $U_{total}$, and polychoric correlation $\rho_{\text{poly(ML)}}$ and $\rho_{\text{poly}}$}
\end{figure}

Figure 2 visualizes the results of the performance evaluation.
From these results, we observe that the polychoric correlation coefficients $\rho_{\mathrm{poly(ML)}}$ and $\rho_{\mathrm{poly}}$ accurately recover the true correlation coefficient $\rho$ for all numbers of categories when $\rho$ is up to approximately $0.9$. 
However, as $\rho$ approaches one, particularly for $\rho$ values of $0.95$ and higher, a deviation from the true $\rho$ emerges and becomes more pronounced as the number of categories increases.
In contrast, the proposed measure exhibits stable performance even for large values of $\rho$, showing behavior consistent with that reported in Section \ref{Sec4.1}.
The results for Cram\'er's coefficient $V^2$ and the total uncertainty measure $U_{\mathrm{total}}$ are also consistent with those reported in Section \ref{Sec4.1}, and therefore are not discussed further here.
The observed difference in performance between the polychoric correlation coefficient and the proposed measure is particularly important in exploratory analyses of large scale data sets, where it is impractical to examine the structure of every contingency table in detail. 
For example, in screening procedures that aim to identify pairs of variables with strong associations, this difference can have a substantial impact on the selected results. 
Taking into account the findings presented above, the proposed measure can be regarded as advantageous for exploratory studies requiring fast and reliable evaluation of association strength.

%%%%%%%%%%%%%%%%%%%%%%%%%%%%%%%%%%%%%%%%%%%
\subsection{Recommendations for selecting parameter $\lambda$ from simulation}\label{Sec4.4}
Based on the findings presented in the previous sections, the recommended range for the parameter $\lambda$ in our proposed framework is $0 \leq \lambda \leq 1$. 
Among these, we particularly advocate the use of $\lambda = 0, 2/3$, and $1$, which correspond respectively to the KL, CR, and Pearson divergences.
The measures $\rho_{(\lambda)}$ associated with these values exhibit distinct properties, as demonstrated in both the simulation studies and the discussion in Section \ref{Sec2.3}, allowing us to formulate concise guidelines for practical use.
For $\lambda = 0$, the measure connects directly to the log–likelihood ratio statistic, which is widely used in tests of independence, and it also coincides with the informational measure by \cite{linfoot1957informational} and Cox–Snell’s $R^2$. 
This provides an interpretability advantage, since the measure offers insight beyond simple association. 
Moreover, as shown in Section \ref{Sec4.1}, it yields values close to the true correlation $\rho$ over the nonnegative range. 
However, compared with $\lambda = 2/3$ and $1$, some settings exhibit noticeable deviations from the $0.95$ coverage probability, which warrants caution.
For $\lambda = 1$, the measure coincides with Pearson's contingency coefficient $C$ and demonstrates relatively stable coverage probabilities across all settings. 
Its drawback, however, is the tendency to underestimate the true correlation $\rho$ over the nonnegative range, as observed in Section \ref{Sec4.1}.
The case $\lambda = 2/3$ does not correspond to a well-known measure or index, yet it shows the second most stable coverage performance (after $\lambda = 1$) and returns values that are close to the true correlation $\rho$.
Taken together, $\lambda = 0$ is recommended when interpretability is the primary concern, $\lambda = 1$ when stability is desired, and $\lambda = 2/3$ when a balance between these properties is preferred.
Although Section \ref{Sec2.2} notes that $\lambda$ can take negative values, including important divergences such as $\lambda = -1$ and $-1/2$, these choices are strongly affected by zero cells and suffer from poor coverage in several settings. 
For these reasons, their use is not recommended.
While sufficiently large sample sizes relative to the number of categories may stabilize the performance of negative $\lambda$ values, simulation results indicate that even in such favorable conditions, $\lambda = 0, 2/3,$ and $1$ continue to outperform them. 
Thus, the use of negative values of $\lambda$ is generally discouraged. 
In the following sections, we therefore focus exclusively on $\lambda$ in $[0, 1]$.

%%%%%%%%%%%%%%%%%%%%%%%%%%%%%%%%%%%%%%%%%%%%%%%%%%%%%%%%%%%%%%%%%%%%%%%%%%%%
\section{Numerical Examples}\label{Sec5}
%%%%%%%%%%%%%%%%%%%%%%%%%%%%%%%%%%%%%%%%%%%%
%\subsection{Health survey data}
We consider the two datasets presented in Tables \ref{Example}(a) and \ref{Example}(b). 
These tables are cross classifications of self-reported mental health status categories and age groups for men and women, obtained from the 1997 National Health Survey conducted in Spain.
For both tables, tests of independence based on Pearson’s chi-square statistic and the log-likelihood ratio statistic yield p-values below $0.001$, indicating the presence of a statistically significant association between the row and column variables.
In this section, we apply the proposed association measures to these data and interpret the degree of association between the two variables as an approximate correlation coefficient. 
By comparing the results obtained from Tables \ref{Example}(a) and \ref{Example}(b), we assess which table exhibits a stronger association.
Following the discussion in Section \ref{Sec4.4}, we adopt the parameter values $\lambda = 0, 2/3,$ and $1$. 
For comparison with the proposed measures, we also consider two divergence-based measures: the total uncertainty coefficient $U_{total}$ and Cram\'er’s coefficient $V^2$.

\begin{table}[ht]\label{Example}
\caption{Self-reported mental health status categories and age groups for men and women, obtained from the 1997 National Health Survey conducted in Spain.}
\centering
\begin{tabular}{c|ccccc|c}
 \multicolumn{7}{c}{(a) Self-Assessment by men}  \\
\hline
 & Very &  &  &  & Very & \\
Age & good & Good & Regular & Bad & bad & Total \\  \hline
16-24 & 145 & 402 & 84 & 5 & 3 & 639 \\
25-34 & 112 & 414 & 74 & 13 & 2 & 615 \\
35-44 & 80 & 331 & 82 & 24 & 4 & 521 \\  
45-54 & 54 & 231 & 102 & 22 & 6 & 415 \\
55-64 & 30 & 219 & 119 & 53 & 12 & 433 \\
65-74 & 18 & 125 & 110 & 35 & 4 & 292 \\
75+ & 9 & 67 & 65 & 25 & 8 & 174 \\  \hline
Total & 448 & 1789 & 636 & 177 & 39 & 3089 \\  \hline 
\end{tabular}

\vskip \baselineskip
\centering
\begin{tabular}{c|ccccc|c}
\multicolumn{7}{c}{(b) Self-Assessment by women}  \\
\hline
 & Very &  &  &  & Very & \\
Age & good & Good & Regular & Bad & bad & Total  \\  \hline
16-24 & 98 & 387 & 83 & 13 & 3 & 584 \\
25-34 & 108 & 395 & 90 & 22 & 4 & 619 \\
35-44 & 67 & 327 & 99 & 17 & 4 & 514 \\  
45-54 & 36 & 238 & 134 & 28 & 10 & 446 \\
55-64 & 23 & 195 & 187 & 53 & 18 & 476 \\
65-74 & 26 & 142 & 174 & 63 & 16 & 421\\
75+ & 11 & 69 & 92 & 41 & 9 & 222 \\  \hline
Total & 369 & 1753 & 859 & 237 & 64 & 3282 \\  \hline
\end{tabular}
\end{table}

Table \ref{Result1} presents the results obtained by applying both the proposed and comparative measures to Tables \ref{Example}(a) and \ref{Example}(b).
The table reports estimated values and $95\%$ confidence intervals for each measure. 
For the comparative methods, only results based on the Simple method are shown, as this is the standard approach typically available.
Interpreted as correlation-like quantities, the results indicate that the proposed measure $\rho_{(\lambda)}$ clearly detects a moderate degree of association between self‐reported mental health status and age group.
For men (Table \ref{Example}(a)), the estimates $\hat{\rho}_{(\lambda)}$ are approximately $0.34$, and the $95\%$ confidence intervals fall within the range of about $0.31$ to $0.37$.
%Interpreted as correlation-like quantities, these values suggest a moderate association, and the minimal variability across different values of $\lambda$ highlights the stability of the proposed estimator.
On the other hand, for women (Table \ref{Example}(b)), the estimates $\hat{\rho}_{(\lambda)}$ are around $0.37$, consistently higher than those for men.
The corresponding confidence intervals, ranging from roughly $0.35$ to $0.41$, also indicate a moderate association.
A noteworthy observation is that, for all choices of $\lambda$, the association is stronger for women than for men. 
This demonstrates the usefulness of the proposed measure for comparing associations across different populations.
However, because the confidence intervals for men and women overlap, the observed difference cannot be regarded as statistically significant despite the higher estimated values for women.
Taken together, the findings based on the proposed measures suggest that self-reported mental health status and age group exhibit a moderate association for both men and women, with women showing a tendency toward a stronger association, although this difference is not statistically significant.
Turning to the comparative measures, their estimated values also indicate a stronger association among women, consistent with the conclusions drawn from the proposed measures.
However, these measures take extremely small values: for men, $U_{total}=0.042$ and $V^2=0.032$; for women, $U_{total}=0.050$ and $V^2=0.039$.
Such small magnitudes are difficult to interpret as effect sizes.
This illustrates a practical advantage of the proposed measure: it provides values that are more interpretable and more directly comparable, thereby offering clearer insight into the strength of association.

\begin{table}[htbp]\label{Result1}
\caption{Results}
\centering
\begin{tabular}{ccccc}
\hline 
 &  & Estimated  & \multicolumn{2}{c}{$95\%$ confidence intervals} \\
Data & Measures & values & by Simple method & by Fisher's $z$ method \\ \hline
Table \ref{Example}(a) 
    & $U_{total}$ & 0.042 & (0.034, 0.050) & -- \\ 
          & $V^2$ & 0.032 & (0.026, 0.038) & -- \\ \cline{2-5}
   & $\rho_{(0)}$ & 0.344 & (0.314, 0.375) & (0.313, 0.375) \\ 
 & $\rho_{(2/3)}$ & 0.340 & (0.310, 0.370) & (0.310, 0.369) \\ 
   & $\rho_{(1)}$ & 0.336 & (0.307, 0.365) & (0.306, 0.364) \\ \hline
Table \ref{Example}(b) 
    & $U_{total}$ & 0.050 & (0.042, 0.059) & -- \\ 
          & $V^2$ & 0.039 & (0.032, 0.045) & -- \\ \cline{2-5}
   & $\rho_{(0)}$ & 0.381 & (0.352, 0.410) & (0.351, 0.410) \\ 
 & $\rho_{(2/3)}$ & 0.374 & (0.347, 0.402) & (0.346, 0.401) \\ 
   & $\rho_{(1)}$ & 0.367 & (0.341, 0.393) & (0.341, 0.392) \\  \hline
\end{tabular}
\end{table}

%%%%%%%%%%%%%%%%%%%%%%%%%%%%%%%%%%%%%%%%%%%%%%%%%%%%%%%%%%%%%%%%%%%%%%%%%%%%
\section{Conclusion}
In this paper, we proposed a family of association measures $\rho_{(\lambda)}$ for two-way contingency tables whose latent distribution can be assumed to be bivariate normal.
The measures are derived by inverting the closed-form approximation between the power-divergence and the latent correlation coefficient established in \citet{urasaki2024generalized}, and can be computed numerically via a simple Newton's method algorithm without evaluating any two-dimensional integrals.
The parameter $-1 \leq \lambda \leq 1$ indexes the family, with specific values recovering well-known special cases: $\lambda = 0$ yields the informational measure of correlation of \citet{linfoot1957informational} and coincides with Cox--Snell's pseudo $R^{2}$ \citep{cox1989analysis}, and $\lambda = 1$ coincides with Pearson's contingency coefficient $C$ \citep{pearson1904theory}.
These connections show that the proposed framework provides a unified generalization of several classical measures.
In addition, we derived asymptotic distributions for $\rho_{(\lambda)}$ and $\rho^{2}_{(\lambda)}$ using the delta method and proposed two methods for constructing confidence intervals: the Simple method, which applies the delta method directly to the estimator, and Fisher's $z$ method, which applies Fisher's $z$-transformation before constructing the interval.
We also introduced detectable correlation thresholds, derived from the chi-squared critical value of the corresponding test of independence, to help practitioners assess whether an observed association is not only statistically significant but also practically meaningful.

Simulation studies showed that the proposed measures approximate the true latent correlation coefficient $\rho$ more faithfully than conventional divergence-based measures such as Cram\'{e}r's coefficient $V^{2}$ and the total uncertainty coefficient, and that they successfully distinguish between weak and moderate associations where existing measures tend to give indistinguishable values.
Coverage probability experiments demonstrated that both confidence interval methods perform well when the sample size is sufficiently large relative to the number of categories.
In sparse settings, however, the normal approximation underlying the interval construction can become unreliable, and caution is required.
Among the parameter values examined, $\lambda = 1$ and $\lambda = 2/3$ consistently showed the most stable coverage, while $\lambda = -1/2$ tended to underperform, particularly when the association was strong.
Furthermore, in a direct comparison with the polychoric correlation coefficient, the proposed measures were computed several thousand times faster for large contingency tables and remained numerically stable even when the latent correlation was close to one, a regime where the polychoric estimator is known to deteriorate.
Although the proposed measures tend to slightly underestimate $\rho$ compared with the polychoric correlation, the degree of underestimation is modest, remaining within approximately 0.01 to 0.03 depending on $\lambda$, and is acceptable for practical purposes.
These findings indicate that the proposed method is particularly advantageous in exploratory analyses where a large number of contingency tables must be screened quickly and reliably.
Based on the simulation results, we recommend the parameter values $\lambda = 0$, $2/3$, and $1$ for practical use.
The choice $\lambda = 0$ is recommended when interpretability is the primary concern, as it connects directly to the log-likelihood ratio statistic and to Cox--Snell's $R^{2}$.
The choice $\lambda = 1$ is recommended when stability of the confidence interval coverage is the main priority, and $\lambda = 2/3$ provides a balance between these two desiderata.
Negative values of $\lambda$, including the important special cases $\lambda = -1$ and $\lambda = -1/2$, are generally not recommended because they are sensitive to zero cells and exhibit poor coverage in several settings.

\begin{comment}
Several directions remain for future work.
The theoretical framework developed here assumes that the latent distribution is bivariate normal, and extending the approach to other latent distributions, such as the bivariate $t$ or skew-normal distribution, would broaden its applicability.
Furthermore, the present study focused on the case of equal marginal probabilities in the simulation experiments; investigating the behavior of the proposed measures under unequal margins would provide additional guidance for practitioners.
Finally, developing bootstrap-based confidence intervals as an alternative to the asymptotic intervals may improve coverage in sparse settings where the normal approximation is unreliable.
\end{comment}

Therefore, the features of the proposed method described above, namely its direct connection to the latent correlation coefficient, its computational efficiency, its numerical stability near the boundary of the parameter space, and its unified inferential framework, make it well-suited for effectively analyzing the strength of association in two-way contingency tables under the bivariate normal assumption. 
We believe that this framework offers new insights into the measurement of association for ordinal categorical data and provides a practical complement to existing methods such as the polychoric correlation coefficient.

\section*{Acknowledgement}
This work was supported by JST SPRING, Grant Number JPMJSP2151.
%Acknowledgements should appear after the body of the paper but before any appendices and be as brief as possible subject to politeness. Information, such as contract numbers, of no interest to readers, should be excluded.

\appendix
%%%%%%%%%%%%%%%%%%%%%%%%%%%%%%%%%%%%%%%%%%%%%%%%%%%%%%%%%%%%%%%%%%%%%%%%%%%%
\section{Monotonicity and uniqueness}\label{A1}
Let
\begin{align*}
S_\lambda(t) =(1-t)^{-\lambda/2}(1-\lambda^2 t)^{-1/2}, \qquad t\in[0,1),
\end{align*}
so that $I_\lambda(t) = \{S_\lambda(t)-1\}/\{\lambda(\lambda+1)\}$.
A direct differentiation yields
\begin{align*}
\frac{\partial}{\partial t}\log S_\lambda(t) = \frac{\lambda}{2(1-t)} + \frac{\lambda^2}{2(1-\lambda^2 t)},
\end{align*}
and hence
\begin{align*}
I_\lambda'(t)
=\frac{S_\lambda(t)}{\lambda(\lambda+1)}
\left\{\frac{\lambda}{2(1-t)} + \frac{\lambda^2}{2(1-\lambda^2 t)}\right\}
=\frac{S_\lambda(t)}{2(\lambda+1)}\left(\frac{1}{1-t} + \frac{\lambda}{1-\lambda^2 t}\right).
\end{align*}
For $-1\leq\lambda\leq1$, we have $1-\lambda^2 t>0$ and $S_\lambda(t)>0$ for all $t\in[0,1)$, so the bracket is strictly positive and consequently $I_\lambda'(t)>0$ on $[0,1)$.
Moreover, $I_\lambda(0)=0$ and $\lim_{t\rightarrow 1} I_\lambda(t)=+\infty$.
Therefore, for each $\hat D_\lambda \geq 0$ there exists a unique solution
$t \in [0,1)$ to \eqref{eq:newton-root}.
We denote this solution by $\hat t_{(\lambda)}$ and define $\hat\rho_{(\lambda)} = \sqrt{\hat t_{(\lambda)}}$.

%%%%%%%%%%%%%%%%%%%%%%%%%%%%%%%%%%%%%%%%%%%%%%%%%%%%%%%%%%%%%%%%%%%%%%%%%%%%
\section{Derivation of initial value}\label{A2}
Considering the binomial expansions of $(1-t)^{-\lambda/2}$ and $(1-\lambda^{2} t)^{-1/2}$ in $S_\lambda(t)$, we obtain
\begin{align*}
S_\lambda(t) &=(1-t)^{-\lambda/2}(1-\lambda^2 t)^{-1/2}\\
&= \left\{1+\frac{\lambda}{2}t + \frac{\lambda(\lambda+2)}{8}t^2 + O(t^3) \right\}\left\{1+\frac{\lambda^2}{2}t + \frac{3\lambda^4}{8}t^2 + O(t^3) \right\} \\
&= 1+\frac{\lambda(\lambda+1)}{2}t + \frac{\lambda(\lambda+1)(3\lambda^2-\lambda+2)}{8}t^2 + O(t^3)
\end{align*}
Therefore, $I_\lambda(t)$ can be rewritten as
\begin{align*}
I_\lambda(t) &= \frac{1}{\lambda(\lambda+1)}\left(S_\lambda(t) - 1\right) \\
&= \frac{1}{\lambda(\lambda+1)}\left\{\frac{\lambda(\lambda+1)}{2}t + \frac{\lambda(\lambda+1)(3\lambda^2-\lambda+2)}{8}t^2 + O(t^3) \right\} \\
&= \frac{1}{2}t + \frac{3\lambda^2-\lambda+2}{8}t^2 + O(t^3)
\end{align*}
for small $t$ we obtain the second-order approximation
\begin{align*}
I_\lambda(t) &\approx \frac{1}{2}t + \frac{3\lambda^2-\lambda+2}{8}t^2.
\end{align*}

Noting that $t \geq 0$, solving this approximation for $t$ yields
\begin{align*}
t = \frac{1}{4c_\lambda}\left\{\left(1+16c_\lambda I_\lambda(t)\right)^{1/2} -1\right\},
\end{align*}
where
\begin{align*}
c_\lambda = \frac{3\lambda^2-\lambda+2}{8}.
\end{align*}
Furthermore, considering the binomial expansion of $\left(1+16c_\lambda I_\lambda(t)\right)^{1/2}$, we may write
\begin{align*}
t = \frac{1}{4c_\lambda}\left\{\left(1+8c_\lambda I_\lambda(t) -32c^2_\lambda I^2_\lambda(t) + O(I^3_\lambda(t))\right) -1\right\}
\end{align*}
Since $I_\lambda(t)$ becomes small when $t$ is small, the approximation 
\begin{align*}
t \approx 2I_\lambda(t) - (3\lambda^2-\lambda+2)I^2_\lambda(t).
\end{align*}
follows.
By substituting $\hat D_\lambda$ for $I_\lambda(t)$, we obtain the initial value
\begin{align*}
\label{eq:initial-value}
t_0 = 2\hat D_\lambda - (3\lambda^2-\lambda+2)\,\hat D_\lambda^2.
\end{align*}

%%%%%%%%%%%%%%%%%%%%%%%%%%%%%%%%%%%%%%%%%%%%%%%%%%%%%%%%%%%%%%%%%%%%%%%%%%%%
\section{Proof of Theorem 3}\label{A3}
Let $D_\lambda(\hat{\bm{p}})$ be the plug-in estimator for the power-divergence $D_\lambda(\bm{p})$.
Then, 
\begin{align*}
\sqrt{n}\left(D_\lambda(\hat{\bm{p}})-D_\lambda(\bm{p})\right) \xrightarrow{d} N\left(0, \sigma^2_{D_{\lambda}} \right),
\end{align*}
where 
\begin{align*}
\sigma^2_{D_{\lambda}} &= \nabla D_\lambda(\bm{p})^T\;(diag(\bm{p})-\bm{p}\bm{p}^T) \; \nabla D_\lambda(\bm{p}), \\
\nabla D_\lambda(\bm{p}) &= \left[\frac{\partial D_\lambda(\bm{p})}{\partial p_{11}}, \frac{\partial D_\lambda(\bm{p})}{\partial p_{12}}, \dots, \frac{\partial D_\lambda(\bm{p})}{\partial p_{rc}} \right], \\
\frac{\partial D_\lambda(\bm{p})}{\partial p_{st}} &= \frac{1}{\lambda}\left(\frac{p_{st}}{p_{s\cdot}p_{\cdot t}}\right)^\lambda - \frac{1}{\lambda+1}\sum^r_{i=1}\frac{p_{it}}{p_{\cdot t}}\left(\frac{p_{it}}{p_{i\cdot}p_{\cdot t}}\right)^\lambda -\frac{1}{\lambda+1}\sum^c_{j=1}\frac{p_{sj}}{p_{s\cdot}}\left(\frac{p_{sj}}{p_{s\cdot}p_{\cdot j}}\right)^\lambda.
\end{align*}
for $s=1, \dots, r$ and $t=1, \dots, c$.

Furthermore, using a first-order Taylor expansion of $I_\lambda (\hat t_{(\lambda)})$ around $t_{(\lambda)}$, we obtain 
\begin{align*}
I_\lambda (\hat t_{(\lambda)}) \approx I_\lambda (t_{(\lambda)}) + I'_\lambda (t_{(\lambda)}) \left(\hat t_{(\lambda)} - t_{(\lambda)}\right).
\end{align*}
For $-1\leq\lambda\leq1$, recall that $t_{(\lambda)}$ and $\hat t_{(\lambda)}$ are implicitly defined by
\begin{align*}
I_\lambda(t_{(\lambda)}) &= D_\lambda(\bm p) \quad \text{and} \quad I_\lambda(\hat t_{(\lambda)})= D_\lambda(\hat{\bm p}).
\end{align*}
Thus, for fixed $\lambda$, we have the approximation
\begin{align*}
D_\lambda(\hat{\bm p}) \approx D_\lambda(\bm p) + I'_\lambda (t_{(\lambda)}) \left(\hat t_{(\lambda)} - t_{(\lambda)}\right).
\end{align*}
From the above approximation we may write, for large $n$,
\begin{align*}
\sqrt{n}\left(\hat t_{(\lambda)} - t_{(\lambda)}\right) \approx \frac{\sqrt{n}\left(D_\lambda(\hat{\bm{p}})-D_\lambda(\bm{p})\right)} {I'_\lambda (t_{(\lambda)})}.
\end{align*}
Hence, by the delta method,
\begin{align*}
\sqrt{n}\left(\hat t_{(\lambda)} - t_{(\lambda)}\right)  \xrightarrow{d} N\left(0, \frac{\sigma^2_{D_{\lambda}}}{\left(I'_\lambda(t_{(\lambda)})\right)^2} \right).
\end{align*}

\clearpage
%%%%%%%%%%%%%%%%%%%%%%%%%%%%%%%%%%%%%%%%%%%%%%%%%%%%%%%%%%%%%%%%%%%%%%%%%%%%
\section{Criteria for assessing association at $\lambda=0$ and $2/3$}\label{A4}
\begin{table}[ht]
\centering
\caption{Thresholds of $\rho_{(0)}$ based on the KL statistic} 
\label{tab:rho_KL}
\begin{tabular}{c|cccccc}
\hline  
& \multicolumn{2}{c}{$n=1000$} & \multicolumn{2}{c}{$n=3000$} & \multicolumn{2}{c}{$n=5000$} \\
df & $\alpha=0.05$ & $\alpha=0.01$ & $\alpha=0.05$ & $\alpha=0.01$ & $\alpha=0.05$ & $\alpha=0.01$ \\ \hline
 1 & 0.0619 & 0.0813 & 0.0358 & 0.0470 & 0.0277 & 0.0364 \\  
 2 & 0.0773 & 0.0957 & 0.0447 & 0.0554 & 0.0346 & 0.0429 \\  
 3 & 0.0882 & 0.1062 & 0.0510 & 0.0614 & 0.0395 & 0.0476 \\  
 4 & 0.0972 & 0.1148 & 0.0562 & 0.0665 & 0.0435 & 0.0515 \\ 
 5 & 0.1049 & 0.1224 & 0.0607 & 0.0708 & 0.0470 & 0.0549 \\ 
 6 & 0.1119 & 0.1291 & 0.0647 & 0.0748 & 0.0502 & 0.0579 \\ 
 7 & 0.1182 & 0.1353 & 0.0684 & 0.0784 & 0.0530 & 0.0607 \\ 
 8 & 0.1240 & 0.1410 & 0.0718 & 0.0817 & 0.0556 & 0.0633 \\ 
 9 & 0.1295 & 0.1464 & 0.0750 & 0.0848 & 0.0581 & 0.0658 \\ 
10 & 0.1347 & 0.1515 & 0.0780 & 0.0878 & 0.0605 & 0.0681 \\ 
15 & 0.1571 & 0.1735 & 0.0911 & 0.1007 & 0.0706 & 0.0781 \\ 
20 & 0.1758 & 0.1920 & 0.1021 & 0.1116 & 0.0791 & 0.0865 \\ 
25 & 0.1922 & 0.2082 & 0.1117 & 0.1211 & 0.0866 & 0.0939 \\  \hline  
\end{tabular}

\vskip\baselineskip
\centering
\caption{Thresholds of $\rho_{(2/3)}$ based on the CR statistic} 
\label{tab:rho_CR}
\begin{tabular}{c|cccccc}
\hline  
& \multicolumn{2}{c}{$n=1000$} & \multicolumn{2}{c}{$n=3000$} & \multicolumn{2}{c}{$n=5000$} \\
df & $\alpha=0.05$ & $\alpha=0.01$ & $\alpha=0.05$ & $\alpha=0.01$ & $\alpha=0.05$ & $\alpha=0.01$ \\ \hline
 1 & 0.0619 & 0.0813 & 0.0358 & 0.0470 & 0.0277 & 0.0364 \\  
 2 & 0.0773 & 0.0957 & 0.0447 & 0.0554 & 0.0346 & 0.0429 \\  
 3 & 0.0882 & 0.1061 & 0.0510 & 0.0614 & 0.0395 & 0.0476 \\  
 4 & 0.0971 & 0.1147 & 0.0562 & 0.0664 & 0.0435 & 0.0515 \\ 
 5 & 0.1048 & 0.1222 & 0.0607 & 0.0708 & 0.0470 & 0.0549 \\ 
 6 & 0.1117 & 0.1289 & 0.0647 & 0.0747 & 0.0501 & 0.0579 \\ 
 7 & 0.1181 & 0.1351 & 0.0684 & 0.0783 & 0.0530 & 0.0607 \\ 
 8 & 0.1239 & 0.1408 & 0.0718 & 0.0817 & 0.0556 & 0.0633 \\ 
 9 & 0.1293 & 0.1461 & 0.0750 & 0.0848 & 0.0581 & 0.0657 \\ 
10 & 0.1345 & 0.1512 & 0.0780 & 0.0877 & 0.0604 & 0.0680 \\ 
15 & 0.1568 & 0.1731 & 0.0910 & 0.1006 & 0.0706 & 0.0780 \\ 
20 & 0.1754 & 0.1914 & 0.1020 & 0.1114 & 0.0791 & 0.0865 \\ 
25 & 0.1916 & 0.2075 & 0.1116 & 0.1209 & 0.0866 & 0.0939 \\  \hline  
\end{tabular}
\end{table}

\newpage

\bibliographystyle{apacite}
\bibliography{paper-ref}

\end{document}